\documentclass[graybox,envcountchap]{svmult}

\usepackage{amsmath}
\usepackage{amssymb}
\usepackage{txfonts}         
\usepackage{color}
\usepackage{helvet}          
\usepackage{courier}         
\usepackage{dirtree}

\usepackage{makeidx}         
\usepackage{graphicx}        
\usepackage{subfig}
\usepackage{microtype}
\usepackage{url}
\usepackage{dsfont}
\usepackage{mathrsfs}

\usepackage{multicol}         
\usepackage[bottom]{footmisc} 

\usepackage[dvipsnames,table]{xcolor}
\usepackage[colorlinks]{hyperref}        
\hypersetup{linkcolor=NavyBlue}
\hypersetup{citecolor=Sepia}
\hypersetup{urlcolor=Sepia}

\usepackage[misc]{ifsym}

\makeindex             

\newcommand{\cc}{{\rm c}}
\newcommand{\dd}{{\rm d}}
\newcommand{\ii}{{\rm i}}
\newcommand{\ee}{{\rm e}}
\newcommand{\hh}{{\rm h}}
\newcommand{\s}{{\rm s}}
\newcommand{\vp}{\varphi}
\newcommand{\vt}{\vartheta}
\newcommand{\pd}{\partial}
\newcommand{\dV}{\dd^4 x \sqrt{-g} \, }
\newcommand{\dVe}{\dd^4 x \sqrt{-g_\ast} \, }
\newcommand{\msun}{{\rm M}_{\odot}}
\newcommand{\pj}{\tilde{p}}
\newcommand{\ej}{\tilde{\varepsilon}}
\newcommand{\pont}{{}^{\ast}RR}

\newcommand{\bl}{\bar{\lambda}}
\newcommand{\bi}{\bar{I}}

\newcommand{\scrp}{\mathscr{P}}
\newcommand{\scrm}{\mathscr{M}}
\newcommand{\scrt}{\mathscr{T}}

\begin{document}

\title*{Neutron stars as extreme gravity probes}
\author{Hector O. Silva}
\institute{Hector O. Silva \at
Max Planck Institute for Gravitational Physics (Albert Einstein Institute), \\
D-14476 Potsdam, Germany \\
\email{hector.silva@aei.mpg.de}}

\maketitle

\abstract{Neutron stars are powerful probes into the extremes of physics. In this
chapter, we will discuss how observations of neutron stars, either in isolation
or in binaries, can be leveraged to test general relativity and constrain competing
theories of gravity.\footnote{Note added in the arXiv version:
this text reflects the subject's status as of January 16th, 2024.}}


\section{Introduction}
\label{sec:intro}

\subsection{Neutron stars, gravity and relativity}
\label{sec:intro_ns_and_rel}

Neutron stars are outstanding laboratories for studying physics at its extreme,
particularly gravity. Let us do a few order-of-magnitude estimates to
understand why this the case.
A typical neutron star has a mass of $M_{\rm NS} = 1.4~\msun$, where $\msun$ is one solar mass
$\msun \approx 1.99 \times 10^{30}$~kg, and a radius $R$ of approximately $12$~km.
We can quantify how relativistic a system is by defining the compactness $C$,
the dimensionless ratio between the system's characteristic mass and size
\begin{equation}
    C = \frac{G M}{R c^2} \,,
\label{eq:compactness}
\end{equation}
where $c = 299,792,458$~m~s$^{-1}$ is the speed of light in vacuum and
$G \approx 6.67 \times 10^{-11}$~${\rm m}^3~{\rm kg}^{-1}~{\rm s}^{-2}$
is Newton's gravitational constant.
For a neutron star, we have $C_\textrm{NS} \approx 0.2$.
To give a sense of what this number means, the Sun's compactness is a minuscule
$C_{\odot} \approx 10^{-6}$, while for a black hole $C_\textrm{BH} = 0.5$, where we took $R$ to be
hole's Schwarzschild radius $R_\textrm{S,\,{\rm BH}} = 2 G M / c^2$.
The Schwarzschild radius of a neutron star is approximately,
\begin{equation}
    R_{\textrm{S,\,NS}} \approx 4.1~{\rm km}\,.
\end{equation}
Comparing this value to their typical $12$~km radius, we find that neutron
stars are at the verge of becoming black holes.
To have a measure of how strong the gravitational field of a neutron star is,
we can use the (Newtonian) escape velocity, $v_\ee$,
\begin{equation}
    v_\ee / c = \sqrt{\frac{2 G M}{R c^2}}  = \sqrt{2 C} \approx 0.6\,.
\end{equation}
That is, the velocity required for a test mass, leaving from the surface of a
neutron star, to escape to infinity is more than half the speed of light.
For all these reasons, black holes, neutron stars (and white dwarfs) are
collectively named ``compact objects.''

These order-of-magnitude estimates show that we cannot make use the
nonrelativistic gravity of Newton to model a neutron star.
Instead, we must use a \emph{relativistic theory of
gravity}~\cite{Oppenheimer:1939ne}.
Einstein's general relativity is the standard theory of relativistic gravity.
In part, this privileged position has been achieved by general relativity's outstanding
success in describing astrophysical phenomena in agreement with observations.
Examples include, in increasing order of how relativistic the system is,
the advance of the periastron of Mercury's orbit around the Sun, the secular
decrease of the orbital period of binary pulsars and the generation and
propagation of gravitational waves produced by coalescing compact binaries.

\subsection{Motivations for testing gravity}

In light of these (and many other) successes, one can ask in all fairness
``\emph{why do we need to test general relativity further?}'' One may answer
this question as follows.

First, it is important to scrutinize general relativity against experiments as
it stands as our best theory to describe one of the fundamental interactions of
nature.
As emphasized by Will~\cite{Will:2014kxa}, the rigidity of general relativity,
in the sense that it has no tunable parameters, makes any new experimental test
(or any refinement of an existing test) a potentially fatal one. In the event
this ever happens, it would necessarily mean the existence of new physics and
force us to reevaluate our understanding of gravity.

Second, from a theoretical perspective, general relativity is an effective
theory of the gravitational interaction that is found to agree with
observations at the energy (curvature) scales that we have probed so far.
It is generally believed that general relativity is corrected at some higher
energy scale. These corrections can be constructed from a ``bottom-up''
approach, by systematically adding to the theory all terms that conform to some
physical principles, or from a ``top-down'' approach, in which such terms would
be determined from the low-energy limit of a more fundamental theory, e.g.,
string theory.
We can then wonder whether these corrections have any observational signatures,
for instance, at the energy scales we can probe with astronomical observations.

Third, there are known examples of theories of gravity which at certain energy
scales are also compatible with observations, whereas at other energy scales predict
deviations of order unity with respect to the predictions of general
relativity.
Hence, whenever a new energy scale becomes experimentally accessible, it is
sensible to be open to the possibility that there may be new gravitational
effects that have so far remained ``screened'' to our observations, and are only
revealed at this new scale.

As we have seen, neutron stars are, second only to black holes, the strongest
sources of gravity we know of. As such, they are natural laboratories to
examine the validity of general relativity in its strong field regime.
Quite remarkably, the earliest work on neutron stars in modified gravity is, to
our knowledge, from 1967~\cite{Salmona:1967zz}. This is the same year in which
Bell Burnell discovered the first pulsar, PSR~B1919$+$21~\cite{Hewish:1968bj}
and Shklovsky correctly interpreted the x-ray source Sco-X1 as being an
accreting neutron star~\cite{Shklovsky:1967}.
We can say, half whimsically, that \emph{the study of neutron stars in modified
theories of gravity is as old as the discovery of neutron stars themselves!}
Quoting Andersson~\cite{Andersson:2017kru}, neutron stars allows us to hear
``whispers from the edge of physics.'' We want to hear, if any, whispers
from new physics beyond general relativity.

\subsection{Goals and the structure of this chapter}

In the remainder of this chapter we will discuss some examples of new phenomena
related to neutron stars that take place when we substitute general relativity
with another gravity theory.
It is not our intent to provide a comprehensive review of the topic, a goal that
Refs.~\cite{Berti:2015itd,Doneva:2017jop,Psaltis:2008bb,Olmo:2019flu} already
accomplish.
In addition, despite their importance in testing gravity, binary pulsars will
not be covered here. The reader can find accounts of this subject in
Ref.~\cite{Wex:2014nva} and in Chapter 3 in this volume by Hu~et~al.~\cite{Hu:2023vsq}.

The approach we adopt here is instead a pedagogical one, where we will go in
depth in some specific examples. We will assume some familiarity with Newtonian
gravity and general relativity at the level of the book by Poisson and Will~\cite{PW:book}.
For the novice, we hope that this chapter can provide a gentle introduction to
the topic and serve as a starting point for further reading or, even,
inspire original research problems.
For the seasoned practitioner, we hope that this chapter contains one or two
new pieces of information, or that a topic familiar to her is presented under a
different light.

This chapter is organized as follows.
In Sec.~\ref{sec:equilibrium} we will review how we obtain neutron star
solutions in general relativity. This will set the stage for us to present how
we obtain neutron star solutions in scalar-tensor gravity, an instructive
example of a modified gravity theory. We will also glance over the properties of
neutron stars in a few other gravity theories.
In Sec.~\ref{sec:tgr_ins} we discuss the challenges in testing gravity with
neutron star observations and discuss some ideas that have been proposed
to circumvent these problems.
Section~\ref{sec:ns_pp} we review the status of x-ray emission from rotating
neutron stars (pulse profile modeling) in modified gravity. This a promising,
but still not fully explored avenue to perform tests of gravity.
In Sec.~\ref{sec:gws}, we review a small selection of tests of gravity using
the observations of gravitational waves signals produced by compact binaries
containing at least one neutron star.
Section~\ref{sec:summary} summarizes this chapter.
We use geometrical units $c = G = 1$ unless stated otherwise, and employ the
mostly plus metric signature.

\section{The equilibrium configuration of a nonrotating neutron star}
\label{sec:equilibrium}

We will start by presenting what modifications to general relativity can cause
on the structure of isolated nonrotating neutron stars.
We begin with a refresher of how we model neutron stars in general relativity,
and then move to extensions thereof.

\subsection{General relativity}
\label{sec:ns_gr}

General relativity is described by the action,
\begin{equation}
    S = \frac{1}{16 \pi} \int \dV R + S_{\rm m}[\Psi_{\rm m}; \, g_{\mu\nu}] \,,
\label{eq:action_gr}
\end{equation}
where $g_{\mu\nu}$ is the metric (with determinant $g$), 
$R$ the Ricci scalar, and $\Psi_{\rm m}$ are the matter fields. Matter fields are coupled
universally to $g_{\mu\nu}$ thus ensuring the weak equivalence principle, that has been
tested to exquisite precision (see, e.g., Ref.~\cite{Will:2014kxa}.)
The field equations of the theory, obtained by varying the action with respect to the inverse metric
$g^{\,\mu\nu}$ (defined such that $g^{\,\mu\alpha} \, g_{\alpha \nu} = \delta^{\,\mu}{}_{\nu}$) are,
\begin{equation}
R_{\mu\nu} - \tfrac{1}{2} R \, g_{\mu\nu} = 8 \pi T_{\mu\nu}\,,
\label{eq:feq_gr}
\end{equation}
where $T_{\mu\nu}$ is the energy-momentum tensor of the matter fields. This
tensor is expressed as a variation of $S_{\rm m}$ with respect to $g^{\,\mu\nu}$
as follows
\begin{equation}
    T_{\mu\nu} = - \frac{2}{\sqrt{-g}} \frac{\delta S_{\rm m}}{\delta g^{\,\mu\nu}} \,.
    \label{eq:tmunu_def}
\end{equation}
The Bianchi identity,
\begin{equation}
    \nabla^{\mu} \left( R_{\mu\nu} - \tfrac{1}{2} R \, g_{\mu\nu} \right) = 0
    \,,
\end{equation}
ensures that $T_{\mu\nu}$ is conserved
\begin{equation}
    \nabla^{\mu} T_{\mu\nu} = 0 \,.
    \label{eq:tmunu_cons}
\end{equation}

Now say that we want to use these field equations to model a nonrotating star.
We will need two ingredients for this. The first one is a choice of the matter content
of the star, while the second one is a choice of the spacetime metric that will
be associated to this matter distribution.
In isolation, the star will be static and spherically symmetric, and we
can use the following line element as an ansatz
\begin{equation}
    \dd s^2 = g_{\mu\nu} \, \dd x^{\,\mu} \, \dd x^{\nu}
    = - e^{2 \nu(r)} \, \dd t^2 + e^{2 \lambda(r)} \, \dd r^2
    + r^2 \, \dd \theta^2 + r^2 \sin^2\theta \, \dd \varphi^2\,.
\label{eq:line_element}
\end{equation}
Here, $\nu$ and $\lambda$ are functions of $r$, called the areal radial
coordinate.
To understand this nomenclature, we can take a constant-in-time
hypersurface (i.e., $\dd t = 0$) and look at a fixed value of $r = R$ (i.e.,
$\dd r = 0$).
We then see from the line element~\eqref{eq:line_element} that $r$ measures
the radius of a 2-sphere.
For reasons that will become clear in a moment, it is convenient to write
\begin{equation}
\lambda = - \tfrac{1}{2} \log[1 - 2 m(r) / r] \,,
\end{equation}
where $m$ is called the mass function. Finally, we will call $\nu$ the lapse
function.\footnote{The reason for this nomenclature is that when we do a $3+1$ decomposition of the spacetime
metric, the $tt$ component is $\alpha^2 - \beta_{k} \, \beta^{k}$, where $\alpha$
is the lapse function and $\beta_{k}$ the shift vector. For $\beta_k = 0$ (called
normal or Eulerian coordinates), we see that $\nu$ is related to the lapse function $\alpha$~\cite{Gourgoulhon:2007ue}.}
As for matter, we will assume that the star can be described by a perfect
fluid. The energy-momentum tensor for a perfect fluid with pressure $p$
and total energy density $\varepsilon$ is
\begin{equation}
T_{\mu\nu} = (\varepsilon + p) u_{\mu} u_{\nu} - p \, g_{\mu\nu} \,,
\label{eq:tmunu_pf}
\end{equation}
where $u^{\,\mu}$ is the four-velocity of an observer comoving with a fluid element.
Because $u^{\,\mu}$ is a timelike vector, it satisfies the normalization $u^{\,\mu}
u_{\mu} = -1$, which combined with Eq.~\eqref{eq:line_element} yields
\begin{equation}
u^{\,\mu} = \left\{ - \exp(-\nu),\, 0,\, 0,\, 0 \right\}.
\end{equation}
The pressure $p$ and energy density $\varepsilon$ are defined with respect to
this observer. The energy density can be decomposed as
$
\varepsilon = \rho \, (1 + \Pi) \,,
$
%
%
where $\rho$ is the baryonic mass density and $\Pi = (\varepsilon - \rho) /
\rho$ is the internal energy per unit baryonic mass.

We can now insert Eqs.~\eqref{eq:line_element}-\eqref{eq:tmunu_pf} in the field equation~\eqref{eq:feq_gr}
and energy-momentum conservation equation~\eqref{eq:tmunu_cons} to obtain the Tolman-Oppenheimer-Volkoff (TOV) equations~\cite{Tolman:1939jz,Oppenheimer:1939ne}:
\begin{subequations}
\label{eq:tov_gr}
\begin{align}
    \frac{\dd m}{\dd r}   &= 4 \pi r^2 \varepsilon \,,
    \label{eq:dmdr_gr}
    \\
    \frac{\dd \nu}{\dd r} &= \frac{1}{r^2}
                             \frac{m + 4 \pi r^3 p}{1 - 2 m/ r} \,,
    \\
    \frac{\dd p}{\dd r}   &= - (\varepsilon + p) \frac{\dd \nu}{\dd r} \,.
\end{align}
\end{subequations}

We see that we have four variables to solve for $(m, \nu, \varepsilon, p)$, but
only three equations.
To close the system, we need to provide an equation of state, that is, a
relationship between the pressure $p$ and energy density $\varepsilon$
\begin{equation}
p = p (\varepsilon) \,.
\label{eq:eos}
\end{equation}
A simple neutron-star equation of state is a polytropic $p = \kappa \, \varepsilon^{1 + 1 / n}$,
where $n = 0.5$ and $\kappa = 2 \times 10^{5}$~km$^{4}$
can be used as representative values~\cite{Kokkotas:2000up}.\footnote{See Tooper~\cite{Tooper:1964,Tooper:1966}
for a detailed discussion relativistic polytropic equations and for the general-relativistic version of the Lane-Emden equations
that describe a polytrope: a self-gravititation body obeying a polytropic equation of state in Newtonian gravity~\cite{Chandrasekhar:1939isss.book}.}
Equations of state often come in tabulated form, and are derived from nuclear
physics calculations that employ a variety of methods and physical assumptions.
See, e.g., Refs.~\cite{Baym:2017whm,Lattimer:2021emm} for reviews articles,
and Refs.~\cite{Typel:2013rza,Oertel:2016bki,CompOSECoreTeam:2022ddl} for an equation-of-state database.
An equation of state of the form~\eqref{eq:eos}
is said to be barotropic.
In general, $p$ could also depend on the temperature $T$, entropy $s$, etc.
However, old neutron stars have temperature well below the Fermi temperature
meaning that in practice they are ``cold'' and can be assumed to be at zero
temperature.\footnote{
In the literature one often finds the terms ``soft'' and ``stiff''
to describe the equation of state. This terminology stems from how fast the
pressure increases as a function of the energy density. The faster this
increase is, the harder it is to compress the fluid and the stiffer the equation
of state is said to be. Hence, stiffer equations of state provide more pressure
against gravitational attraction and can support stars with larger masses.
Conversely, softer equations result in stars with smaller radii.}

The star's radius $R$ is the location in which the pressure $p$
vanishes. The star's gravitational mass then follows from Eq.~\eqref{eq:dmdr_gr}
\begin{equation}
    M = m(R) = \int_{0}^{\,R} 4 \pi r^{2} \varepsilon \, \dd r \,,
\end{equation}
which justifies the naming $m(r)$ as the mass function.
The upper integration limit follows from $\varepsilon$ being zero in vacuum; see~Eq.~\eqref{eq:dmdr_gr}.

It is also useful to define the star's baryonic mass $M_{\rm b}$ as an integral of
the baryonic mass density $\rho$ in the spatial volume occupied by the star,
\begin{equation}
    M_{\rm b} = 4 \pi m_{\rm b} \int_{0}^{\,R}  \frac{n_{\rm b} r^2}{\sqrt{1 - 2 m / r}} \, \dd r \,.
\end{equation}
Here, we wrote $\rho = m_{\rm b} n_{\rm b}$, where $m_{\rm b} \approx 1.66
\times 10^{-24}$~g is the atomic mass unit and $n_{\rm b}$ the number density
of baryons.
From $M$ and $M_{\rm b}$ we can define the normalized (gravitational) binding energy $E_{\rm b}$,
\begin{equation}
    E_{\rm b} = (M_{\rm b} - M) / M \,,
\label{eq:def_binding_energy}
\end{equation}
which is positive for a bound system as a star. Typical values of $E_{\rm b}$
for neutron stars range between 0.1 and 0.2, approximately.

An important result for us is the Jebsen-Birkhoff theorem.
%
This theorem states that the \emph{vacuum exterior} spacetime of any, possibly
time dependent, spherically symmetric mass configuration, is given by the
Schwarzschild metric.
Thus, the spacetime outside the star has line element
\begin{equation}
    \dd s^2 =
    - \, (1 - 2 M / r) \, \dd t^2
    + (1 - 2 M /r)^{-1} \, \dd r^2
    + r^2 \, \dd \theta^2
    + r^2 \sin^2\theta \, \dd \varphi^2
    \,,
\label{eq:line_element_sch}
\end{equation}
where $M = m(R)$.
If one is interested only in studying the exterior spacetime of the
neutron star, one can also work directly with Eq.~\eqref{eq:line_element_sch},
using values of $M$ and $R$ that are typical of neutron stars.

We now have all the ingredients to obtain neutron star solutions in
general relativity.
To solve the TOV equations we would in principle start
from the star's center $r = 0$ and proceed outwards until we locate the star's radius.
In practice, however, we find that the TOV equations are singular at $r = 0$.
Instead,we use an approximate, power-series solution of the relevant variables around $r=0$.
For instance, the first terms in the series expansion are:
\begin{subequations}
\label{eq:near_core_series}
\begin{align}
    m &= \tfrac{4}{3} \pi \varepsilon_\cc r^3 + {\cal O}(r^5) \,,
    \\
    p &= p_\cc - \tfrac{2}{3} \pi (\varepsilon_\cc + p_\cc) (\varepsilon_\cc + 3p_\cc) r^2 + {\cal O}(r^4)\,,
    \\
    \nu &= \nu_\cc + \tfrac{2}{3} \pi (\varepsilon_\cc + 3p_\cc) r^2  + {\cal O}(r^4)\,,
\end{align}
\end{subequations}
where $\nu_\cc$ is a free constant.
The integration of the TOV equations~\eqref{eq:tov_gr} then proceed as follows:

\begin{svgraybox}
\textbf{Box 1}: TOV integration in general relativity
\begin{enumerate}
    \item Choose a value of the energy density $\varepsilon_{\cc}$ at the star's center $r = r_{\cc}$.
    The central pressure $p_{\cc} = p(\varepsilon_{\cc})$ is fixed by the equation of state~\eqref{eq:eos}.
    \item Integrate the TOV equations~\eqref{eq:tov_gr} outwards starting from $r = r_\cc$, with initial conditions
    for $m$, $p$ and $\nu$ given by Eqs.~\eqref{eq:near_core_series} evaluated at $r = r_\cc$.
    \item Proceed with the integration until the pressure $p$ vanishes. This fixes the radius $R$ and the mass
    $M = m(R)$.
    \item Fix the constant $\nu_\cc$, such that $\nu$ matches smoothly $\exp(2\nu(R)) = 1 - 2 M /R$ at $r=R$.
\end{enumerate}
\label{box:gr_algo}
\end{svgraybox}

If we repeat these steps for a range of values of $\varepsilon_\cc$ we
obtain a one-parameter sequence of solutions.
These solutions can be plotted in a $MR$-plane, producing what is known as the
mass-radius relation.

\begin{figure}[t]
\centering
\includegraphics[width=0.75\columnwidth]{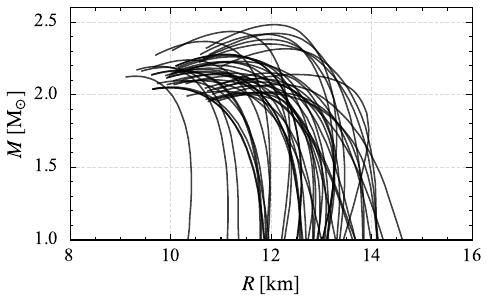}
\caption{Neutron star mass-radius relations obtained by integrating the TOV equation for
a large sample of equations of state.}
\label{fig:mass_radius}
\end{figure}

In Fig.~\ref{fig:mass_radius}, we show several of such curves for a selection of equations of
state that support neutron stars with masses $M \geqslant 2~\msun$, as required by the precise
inferred masses of pulsars
J0348$+$0432, $M = 2.01 \pm 0.04~\msun$~\cite{Antoniadis:2013pzd},
and J0740$+$6620, $M = 2.08\pm0.07~\msun$~\cite{NANOGrav:2019jur,Fonseca:2021wxt}, both at $1\sigma$ level.
The equations of state we used form a subset of those referred to in Ref.~\cite{Silva:2020acr}.
The large variability in the predicted masses and radii is a consequence of our
current poor understanding of the equation of state at high densities.

\subsection{Bergmann-Wagoner scalar-tensor gravity}
\label{sec:ns_stt}

\subsubsection{Motivation}

In general, the steps to obtain neutron star solutions taken previously can be
repeated in any other gravity theory with some modifications.
As a simple, but nontrivial example, we will consider the Bergmann-Wagoner
scalar-tensor gravity~\cite{Bergmann:1968ve,Wagoner:1970vr}.
This is arguably the simplest extension to general relativity, and it is, to
our knowledge, the earliest beyond-general-relativity theory in which neutron
stars were studied~\cite{Salmona:1967zz}.

\subsubsection{Theoretical minimum}

The action of this theory is given by,
\begin{equation}
    S = \frac{1}{16 \pi G_{\ast}}
    \int \dVe \left[ R_{\ast} - 2 (\partial \varphi)^2 - 4 V(\varphi) \right]
    + S_{\rm m} \left[\Psi_{\rm m}; \, A^{2}(\varphi) \, g_{\ast\mu\nu} \right] \,,
\label{eq:action_stt}
\end{equation}
where $\vp$ is scalar field with canonical kinetic term $(\pd \vp)^2 = \pd_{\alpha}\vp \, \pd^{\alpha}\vp$,
and self-interaction potential $V$. The spacetime metric is $g_{\ast\mu\nu}$ with respect to which
we define the Ricci scalar $R_{\ast}$, as usual. Lastly, $A$ is a nonvanishing a function of $\vp$.

Why are we introducing asterisks in the metric and Ricci scalar here?
And why did we reinstate the gravitational constant $G_{\ast}$? [Compare against Eq.~\eqref{eq:action_gr}.]
The reason is that in scalar-tensor theories we have the freedom to perform
field redefinitions and metric rescalings to write the action in different
forms, known as ``frames.''

Equation~\eqref{eq:action_stt} describes the theory in the
\emph{Einstein frame} characterized by the absence of a coupling between $\vp$ and the Ricci
scalar $R_{\ast}$.
The main difference between Eq.~\eqref{eq:action_stt} and the action describing a scalar field
minimally coupled to gravity is in the universal coupling between all matter
fields $\Psi_{\rm m}$ with the metric
\begin{equation}
    \tilde{g}_{\mu\nu} = A^2(\vp) \, g_{\ast\mu\nu}\,.
    \label{eq:jordan_to_einstein}
\end{equation}
Hence, the theory obeys the weak equivalence principle, and massive test particles
follow geodesics in $\tilde{g}_{\mu\nu}$. Massless particles, such as photons, follow
null geodesics which are invariant under the transformation~\eqref{eq:jordan_to_einstein},
sometimes called a conformal transformation.
We call $\tilde{g}_{\mu\nu}$ the \emph{Jordan frame} metric. The reason for this nomenclature
is that we can rewrite Eq.~\eqref{eq:action_stt} replacing $g_{\ast}$ and $R_{\ast}$ by their
counterparts calculated with respect to $\tilde{g}_{\mu\nu}$.
We use tildes and asterisks to denote quantities in the Jordan and Einstein
frames respectively.
The advantage of working in the Einstein frame is that the field equations
are considerably simpler.

In the following, we will assume that the potential $V$ is zero, i.e., the
scalar field is massless and non-self-interacting.
Performing the variation of the action~\eqref{eq:action_stt} with respect to
$g^{\,\mu\nu}_{\ast}$ and $\vp$ gives us the field equations,
\begin{subequations}
\label{eq:eom_stt}
\begin{align}
    & R_{\ast\mu\nu} - \tfrac{1}{2} R_{\ast} \, g_{\ast\mu\nu} = 2 \pd_{\mu}\vp \, \pd_{\nu}\vp - g_{\ast\mu\nu} (\pd \vp)^2 + 8 \pi T_\ast \,,
    \label{eq:eom_g_stt}
\\
    & \Box_{\ast} \vp = - 4 \pi \alpha T_{\ast} \,,
    \label{eq:eom_phi_stt}
\end{align}
\end{subequations}
where $\Box_{\ast} = g_{\ast\mu\nu} \nabla_{\ast}^{\mu} \nabla_{\ast}^{\nu}$ is the d'Alembertian operator,
$T_{\ast} = T_{\ast}^{\mu\nu} g_{\ast\mu\nu}$ is the trace of the energy-momentum tensor,
which is defined in analogy with Eq.~\eqref{eq:tmunu_def},
\begin{equation}
    T_{\ast\mu\nu} = - \frac{2}{\sqrt{-g_\ast}} \frac{\delta S_{\rm m}}{\delta g_{\ast}^{\,\mu\nu}} \,,
\end{equation}
and
\begin{equation}
\alpha(\vp) = \frac{\dd}{\dd \vp} \log A(\vp) \,.
\label{eq:stt_alpha}
\end{equation}
We call $\alpha$ the scalar-field-matter coupling as it controls how
strongly matter sources $\vp$ in Eq.~\eqref{eq:eom_phi_stt}. We have also set $G_{\ast} = 1$.

We define the Jordan frame energy-momentum tensor
$\tilde{T}^{\mu\nu}$ in analogy with Eq.~\eqref{eq:tmunu_def},
\begin{equation}
    \tilde{T}_{\mu\nu} = - \frac{2}{\sqrt{-\tilde{g}}} \frac{\delta S_{\rm m}}{\delta \tilde{g}^{\,\mu\nu}} \,.
\end{equation}
From their definitions, we see that the Einstein and Jordan frames
energy-momentum tensors satisfy the following relations
\begin{equation}
T_{\ast}^{\mu\nu} = A^6(\vp) \tilde{T}^{\mu\nu},
\quad
T_{\ast\mu\nu} = A^2(\vp) \tilde{T}_{\mu\nu},
\quad
\textrm{and}
\quad
T_{\ast} = A^4(\vp) \tilde{T} \,,
\end{equation}
and their divergences are
\begin{equation}
    \tilde{\nabla}^{\mu} \tilde{T}_{\mu\nu} = 0 \,,
    \quad
    \textrm{and}
    \quad
    \nabla_{\ast}^{\mu} T_{\ast\mu\nu} = \alpha(\vp) \, T_{\ast} \, \nabla_{\ast\nu} \vp \,.
\end{equation}
As already mentioned, this shows that test particles follow geodesics in the
Jordan, but not in the Einstein frame.

The physically measured pressure and densities of a fluid are those in the
Jordan frame to which all matter (including our measurement devices)
are universally coupled to.
In analogy with Eq.~\eqref{eq:tmunu_pf}, we write for a perfect fluid:
\begin{equation}
    T_{\ast\mu\nu} = (\varepsilon_{\ast} + p_{\ast}) u_{\ast\mu} u_{\ast\nu} - p_{\ast} \, g_{\ast\mu\nu} \,.
\end{equation}
Taking the trace, we obtain,
\begin{equation}
    T_{\ast} = T_{\ast\mu\nu}\, g_{\ast}^{\mu\nu} = 3 p_{\ast} - \varepsilon_{\ast}  = A^4(\vp) \, ( 3 \pj - \ej )\,,
\end{equation}
which is nonzero in general.\footnote{Note that $T_{\ast} = \tilde{T} = 0$ if $\pj = \ej /
3$. This corresponds to conformal fluids and they are interesting in the context
of the gravity-fluid correspondence~\cite{Rangamani:2009xk}. Conformal fluids
have a sound speed squared $v_{\rm s}^2 = {\dd p / \dd \varepsilon} =
1/3$ that was thought to be an upper bound on the sound speed of
any medium. This bound has been challenged by the observation of
approximately $2~\msun$ neutron stars. See, for instance,
Refs.~\cite{Alsing:2017bbc,Bedaque:2014sqa} for a discussion.}
Thus, stars can act as scalar field sources; see~Eq.~\eqref{eq:eom_phi_stt}.

The limit of general relativity (plus a minimally coupled scalar field)
is obtained when we take $\alpha$ to be zero. In this limit, the Jordan- and
Einstein-frame metrics coincide and we can safely drop their identifiers.

\subsubsection{The equations of hydrostatic equilibrium}

The generalization of the TOV equations can be obtained in the same way as in
general relativity.
We assume the line element:
\begin{equation}
    \dd s_{\ast}^2 = g_{\ast\mu\nu} \, \dd x^{\,\mu} \, \dd x^{\nu}
    = - e^{2 \nu(r)} \, \dd t^2 + e^{2 \lambda(r)} \, \dd r^2
    + r^2 \, \dd \theta^2 + r^2 \sin^2\theta \, \dd \varphi^2\,,
\label{eq:line_element_eins}
\end{equation}
and a Jordan-frame perfect fluid described by
\begin{equation}
    \tilde{T}_{\mu\nu} = (\ej + \pj) \tilde{u}_{\mu} \tilde{u}_{\nu} - \pj \, \tilde{g}_{\mu\nu} \,,
\label{eq:tmunu_pfj}
\end{equation}
where $\tilde{u}_{\mu} \tilde{u}^{\,\mu} = -1$ and hence $\tilde{u}^{\,\mu} =
A^{-1}(\vp) \, \{ -\exp(-\nu),\, 0,\, 0,\, 0 \}$.

Using Eqs.~\eqref{eq:line_element_eins} and~\eqref{eq:tmunu_pf}, we find from
the field equations~\eqref{eq:eom_stt} the TOV equations in scalar-tensor gravity
\begin{subequations}
\label{eq:tov_stt}
\begin{align}
    \frac{\dd m}{\dd r} &= 4 \pi r^2 A^4 \ej
    + \frac{1}{2} r (r - 2 m) \psi^2,
    \label{eq:dmdr_stt}
    \\
    \frac{\dd \nu}{\dd r} &= 4 \pi r^2 A^4 \frac{\pj}{r - 2 m} + \frac{1}{2} r \psi^2 + \frac{m}{r(r - 2m)},
    \\
    \frac{\dd \pj}{\dd r} &= - (\pj + \ej)
    \left( \frac{\dd \nu}{\dd r} + \alpha \psi \right),
    \\
    \frac{\dd\psi}{\dd r} &=
    4 \pi \frac{r A^4}{r - 2m} [\alpha(\ej - 3\pj) + r \psi(\ej - \pj)]
    - \frac{2 (r - m)}{r (r - 2m)} \psi,
    \label{eq:dpsidr_stt}
\end{align}
\end{subequations}
where we defined $\psi = \dd \vp / \dd r$, and Eq.~\eqref{eq:dpsidr_stt} follows from~\eqref{eq:eom_phi_stt}.

The main difference with respect to general relativity is the presence of the
scalar field, and this will force us to introduce some modifications to the
integration procedure to obtain neutron star solutions and determine their properties.
To understand why this is the case, we evaluate Eqs.~\eqref{eq:dmdr_stt}
and \eqref{eq:dpsidr_stt} outside the star by setting $\ej = \pj = 0$:
\begin{subequations}
\label{eq:tov_stt_vac}
\begin{align}
\frac{\dd m}{\dd r} &=
\frac{1}{2} r (r - 2 m) \psi^2,
\\
\frac{\dd \nu}{\dd r} &=
\frac{1}{2} r \psi^2 + \frac{m}{r(r-2M)},
\\
\frac{\dd\psi}{\dd r} &=
- \frac{2 (r - m)}{r (r - 2m)} \psi.
\end{align}
\end{subequations}
We see that unless the scalar field is constant, $\psi = 0$, the mass function $m$ has a nonzero gradient
in vacuum and the lapse behaves differently with respect to general relativity.
These facts have two important consequences.
The first is that the mass function evaluated at $r = R$ does not determine the
star's total mass anymore.
The second is that the spacetime in the star's exterior is no longer
Schwarzschild.

To determine the star's mass, we can expand Eqs.~\eqref{eq:tov_stt_vac} in
a power series in $1/r$ and take the far-field limit $R / r \ll 1$. The result is
\begin{subequations}
\label{eq:stt_vac_farfield}
\begin{align}
    m &= M - q^2 / (2r) - q^2 M / (2 r^2) + O(r^{-3}) \,,
    \\
    e^{2 \nu} &= 1 - 2 M /r  + O(r^{-3}) \,,
    \\
    \vp &= \vp_{0} + q / r + q M / r^2 + O(r^{-3}) \,,
\end{align}
\end{subequations}
where we have identified the $1/r$ coefficient in the lapse's expansion as the mass $M$,
and $\vp_{0}$ and $q$ are constants. The constant $\vp_{0}$ can be
interpreted as the ambient scalar field in which the
star is embedded, while $q$, in analogy with electrostatics, is called the
(monopole) scalar charge. This charge is not a conserved charge as
in electromagnetism.
For a constant scalar field the charge vanishes and then $\vp = \vp_{0}$, $m = M$ and $\exp(2\nu) = 1 - 2M/r$.
Hence, we recover the Schwarzschild exterior solution.

Equations~\eqref{eq:stt_vac_farfield} suggest that the vacuum spacetime of a neutron star
in scalar-tensor gravity cannot be obtained in closed form, contrary to general relativity.
It turns out that this is possible if we adopt a different coordinate system,
named after
Just~\cite{Buchdahl:1959nk,Coquereaux:1990qs,Damour:1992we,Just:1959}.
The line element is:
\begin{equation}
    \dd s_{\ast}^2 = - f^{b/a} \, \dd t^2 + f^{-b/a} \, \dd \varrho^2
    + \varrho^2 f^{1 - b/a} \, \dd \theta^2
    + \varrho^2 f^{1 - b/a} \, \sin^2 \theta \, \dd \phi^2 \,,
    \label{eq:line_element_just}
\end{equation}
where $b = 2 M$ and $f = 1 - a / \varrho$. The constant $a$ has units of length
and is related to the star's mass $M$ and the scalar charge $q$, as we will
see shortly.
The Just and Schwarzschild radial coordinates are related as
\begin{equation}
r = \varrho \, (1 - a/\varrho)^{(1 - b/a)/2} \,,
\end{equation}
which cannot be inverted analytically. However, $r \simeq \varrho$ when $a / \varrho \ll 1$,
for instance, far away from the star.

Using Eq.~\eqref{eq:line_element_just}, we can solve analytically Eq.~\eqref{eq:eom_phi_stt}
in the exterior region of a star:
\begin{equation}
\vp = \vp_{0} + (q/a) \log(1 - a/\varrho) \,.
\label{eq:phi_just}
\end{equation}
In the large $\varrho$-limit, $\vp \simeq \vp_{0} + q / \varrho \simeq \vp_0 + q / r$,
confirming that  $q$ is the scalar charge.
The constants $a$, $b$ and $q$ are not independent, and satisfy the constraint
\begin{equation}
a^2 - b^2 - (2q)^2 = 0 \,,
\label{eq:just_constraint}
\end{equation}
that follows from the $\theta\theta$ component of Eq.~\eqref{eq:eom_g_stt} in vacuum.
From Eq.~\eqref{eq:just_constraint}, we can express the ratio $a/b$ appearing in Eq.~\eqref{eq:line_element_just} as
\begin{equation}
    a/b = \sqrt{1 + Q^2}\,,
\end{equation}
where we introduced the dimensionless scalar-charge-to-mass ratio
\begin{equation}
Q = q / M \,.
\label{eq:def_Q}
\end{equation}
Therefore, the exterior metric of a neutron star in scalar-tensor theory is fully determined
by $M$ and $Q$. The constant $Q$ controls how much the spacetime deviates from the Schwarzschild
spacetime, which is recovered for $Q = 0$.

The benefit of having this analytical solution is that we can now integrate the
TOV equations~\eqref{eq:tov_stt} only within the star, and determine $M$,
$\vp_{\infty}$ and $q$ by imposing that the variables match smoothly
Eqs.~\eqref{eq:line_element_just} and~\eqref{eq:phi_just} at the star's radius.
This makes the problem of finding neutron star solutions similar to that in
general relativity.
This calculation was first done in Ref.~\cite{Damour:1993hw} (see also Ref.~\cite{Creci:2023cfx}) and yields:
\begin{subequations}
\label{eq:matching_just}
\begin{align}
    q &= - M \psi_\s / \nu'_\s \,,
    \\
    \vp_{0} &= \vp_{\rm s} - (q/M) \nu_{\rm s} = \vp_{\s}
    + \frac{\psi_\s}{ (\nu'^2_\s + \psi_\s^2)^{1/2} }
    {\rm arctanh}\left[
    \frac{(\nu'^2_\s + \psi_\s^2)^{1/2}}{\nu'_\s + 1 / R}
    \right] \,,
    \label{eq:vp0}
    \\
    M &=
    {R^2 \nu'_\s}
    \left( 1 - \frac{2 m_\s}{R} \right)^{1/2}
\exp\left\{
    - \frac{\nu'_\s}{ (\nu'^2_\s + \psi_\s^2)^{1/2} }
    {\rm arctanh}\left[
    \frac{(\nu'^2_\s + \psi_\s^2)^{1/2}}{\nu'_\s + 1 / R}
    \right]
\right\} \,,
\end{align}
\end{subequations}
where a prime indicates a derivative with respect to $r$ and the subscript
``s'' means that a variable is evaluated at the star's surface $r = R$.

We now have all the ingredients to obtain neutron star solutions in scalar-tensor gravity.
To solve the TOV equations~\eqref{eq:tov_stt} we proceed as follows:
\begin{svgraybox}
    \textbf{Box 2}: TOV integration in scalar-tensor gravity given a conformal factor $A(\vp)$
    \begin{enumerate}
        \item Choose a value of the energy density $\tilde{\varepsilon}_{\cc}$ at the star's center $r = r_{\cc}$.
            The central pressure $\tilde{p}_{\cc} = \tilde{p}(\tilde{\varepsilon}_{\cc})$ is fixed by the equation of state~\eqref{eq:eos}.
        \label{itm:stt_int_ini}
        \item Integrate the TOV equations~\eqref{eq:tov_stt} outwards starting
        from $r = r_\cc$, with initial conditions $\vp_\cc = \vp(r_\cc)$,
        $\psi_\cc = 0$, $\nu_\cc = \nu(r_\cc)$ (a freely chosen constant),
        and $m_\cc = 0$ for the scalar field, its derivative, lapse
        and mass functions.
        \item Proceed with the integration until the pressure $p$ vanishes. This fixes the radius $R$. From
        the values of $\nu'_\s$, $\psi_\s$, and $m_\s$ determine $q$, $\vp_{0}$ and $M$ using Eqs.~\eqref{eq:matching_just}.
    \item Fix the constant $\nu_\cc$, such that $\nu$ matches smoothly $\nu_{\rm s}$ defined in the second equality in Eq.~\eqref{eq:vp0}.
        \label{itm:stt_int_fin}
\label{box:stt_algo}
\end{enumerate}

    Often one requires that $\vp_{0}$ has a specific value fixed e.g.,
    either by observational constraints or by the scalar field's cosmological
    evolution. In this case, $\vp_\cc$ \emph{is not a free constant}. Instead, we must apply a root-finding
    algorithm to determine $\vp_\cc$ such that the outcome of steps \ref{itm:stt_int_ini} to \ref{itm:stt_int_fin} is $\vp_0$ with the targeted value.
\end{svgraybox}

We stress that one can also numerically integrate the TOV
equations~\eqref{eq:tov_stt_vac} in vacuum up to a large $r = r_{\rm max}$.
We then fix the three constants $\vp_{0}$, $q$ and $M$, from the knowledge of
$m$, $\vp$ and $\psi$ using the three equations~\eqref{eq:stt_vac_farfield} evaluated at $r_{\rm max}$.
The advantage of this approach is that it generalizes to other theories
of gravity in which an exact exterior spacetime solution does not exist.
Of course, one has to first obtain the equivalent of
Eqs.~\eqref{eq:stt_vac_farfield} in the theory of interest.
This method gives the same results as the approach presented in Box~2.

Finally, we emphasize that $q$, $M$ and $R$ are computed in the Einstein frame.
From $\dd\tilde{s}^2 = A^2(\vp) \, \dd s_{\ast}^2$, it is easy to see that the
Jordan-frame radius $\tilde{R}$ is:
\begin{equation}
    \tilde{R} = A^{2}(\vp_{\rm s}) \, R\,,
\end{equation}
whereas a translation from Einstein to Jordan frame of Eqs.~\eqref{eq:stt_vac_farfield} shows
that the Jordan-frame gravitational mass $\tilde{M}$ and scalar charge $\tilde{q}$ are:
\begin{equation}
\tilde{M} = M \, \frac{1}{A_0} \, \frac{ 1 - Q \, \alpha_0 }{1 + \alpha^2_0} \,,
\quad
\tilde{q} = q \, \frac{\alpha_0}{A^3_0} \frac{2}{1 + \alpha^2_0} \,,
\end{equation}
respectively, and $A_0 = A(\vp_0)$ and $\alpha_0 = \alpha(\vp_0)$.
Details of this calculation can be found in Refs.~\cite{Creci:2023cfx,Damour:1992we,Pani:2014jra}.
In scalar-tensor theories, it is the Einstein-frame mass that has desired
physical properties. See Ref.~\cite{Doneva:2022ewd}, Sec.~III.A and references
therein.

\subsubsection{Spontaneous scalarization}
\label{sec:neutron_star_scalarization}

The attentive reader may at this point start becoming anxious about the fact
that the function $A(\vp)$ has been left undetermined so far. What functional
form do we chose for it?
Let us Taylor-expand $\alpha(\varphi)$ [recall~Eq.~\eqref{eq:stt_alpha}] as
follows
\begin{equation}
    \alpha(\vp) = \alpha_0 + \beta_0 (\varphi - \varphi_0) + \dots \,,
    \quad
    \beta_0 = (\dd \alpha / \dd \vp)_{\vp = \vp_0} \,,
\end{equation}
which we truncate at linear order in $\vp - \vp_{0}$.
One can show that the deviations from general relativity in the post-Newtonian
approximation are proportional to the square of $\alpha_0$.
This term is related to the constant $\omega_{\rm BD}$ in Brans-Dicke
theory~\cite{Brans:1961sx}, and is constrained by Solar System test to be
smaller than $3.5 \cdot 10^{-3}$.
For this reason, it is tempting to imagine that higher-order powers in this
series will contribute very little to the theory's prediction.
This is not the case, however, as first observed by Damour and
Esposito-Far\`ese~\cite{Damour:1993hw,Damour:1996ke}.

Consider for the moment that the value of the scalar field is so small, that
we can treat it as a small perturbation $\delta \vp$ around a general-relativistic
background. Let us write $\vp = \vp_0 + \delta \vp$, set without loss of
generality $\vp_0 = 0$, and $\alpha_0 = 0$.
By linearizing Eq.~\eqref{eq:eom_phi_stt} in $\delta \vp$, we find
\begin{equation}
    \Box_{\ast} \delta \vp = - 4 \pi \beta_0 \tilde{T}\delta \vp \,,
    \label{eq:box_linearized_def}
\end{equation}
where the conformal factor $A \approx 1$ in our approximation and hence $\tilde{T} \simeq T_{\ast} = T$.
This equation has the same form of a massive Klein-Gordon equation,
\begin{equation}
    \left[ \Box_{\ast} - \mu^2_{\rm eff}(r) \right] \delta \vp = 0 \,, \quad \mu^2_{\rm eff} = - 4 \pi \beta_0 T \,,
\label{eq:linearized_def}
\end{equation}
where $\mu_{\rm eff}$ is a position-dependent effective mass for the scalar field.

There are two important observations we can make from this equation.
First, $\delta \vp = 0$ is a solution. Second, as long as $\beta_0 T > 0$,
the perturbations can have an imaginary effective mass.
This means that some modes of $\delta \vp$ can become tachyonic unstable,
causing $\delta \vp$ to exponentially grow in time; see Ref.~\cite{Doneva:2022ewd},
Sec.~II, for a review.
In general, the trace $T$ is negative, meaning that we need $\beta_0$ negative for this
to happen.

To find the threshold for this instability, we can first exploit the
spherical symmetry of the background spacetime and do a harmonic mode
decomposition as follows:
\begin{equation}
    \delta \vp(t,r,\theta,\vp) = \sum_{l,\, m}
    \frac{\psi(r)}{r}
    Y_{l m} (\theta, \vp)
    \, \ee^{\ii \sigma t}
    \,,
    \label{eq:scalarization_ansatz}
\end{equation}
where $Y_{lm}$ are the spherical harmonics and $\sigma$ is a frequency.
Substituting Eq.~\eqref{eq:scalarization_ansatz} in
Eq.~\eqref{eq:linearized_def} and performing separation of variables, we
obtain an ordinary differential equation for $\psi$.
This equation can be brought to the form of a Schr\"odinger-like equation
\begin{equation}
    \left[ \frac{\dd^2}{\dd x^2} + \sigma^2 - V_{\rm eff}(x) \right] \, \psi = 0\,,
\label{eq:schrodinger_like_eq}
\end{equation}
by introducing a new radial coordinate coordinate $x$, related to $r$ as
\begin{equation}
    \frac{\dd x}{\dd r} = (1 - 2 m / r)^{-1/2} \, \ee^{-\nu} \,.
\end{equation}
Here, $\nu$ and $m$ are the lapse and mass functions of the background spacetime, respectively.
In Eq.~\eqref{eq:schrodinger_like_eq}, we identified an effective potential
$V_{\rm eff}$ for $\psi$, namely,
\begin{equation}
    V_{\rm eff} = \ee^{2 \nu} \left[
        \frac{l(l+1)}{r^2} + \frac{2 m}{r^3} + 4 \pi (p - \varepsilon) + \mu^2_{\rm eff}
    \right] \,.
    \label{eq:effective_potential}
\end{equation}
We can now identify the threshold for the instability by searching for $\sigma = 0$ (zero mode
or bound state) solutions of Eq.~\eqref{eq:schrodinger_like_eq}.
Bound state solutions
are characterized by being regular at the center of the star ($r = 0$) and for approaching a constant
(in our case chosen to be zero) at spatial infinity.\footnote{The black-hole perturbation theory
enthusiast will note that in vacuum, the effective potential~\eqref{eq:effective_potential} reduces to that of massless scalar
perturbations in a Schwarzschild spacetime, as expected. See, for instance,
Chapter 9, Sec 9.2.2, in this volume~\cite{Franchini:2023eda}.}
This constitutes a boundary-value problem, where given a spacetime background
and a value of $l$, we try to find a value of $\beta_0$ for which a solution
$\psi$ that satisfies these boundary conditions exists.
In practice, we can solve this boundary-value problem using a shooting method.

\begin{figure}[t]
\centering
\includegraphics[width=0.75\columnwidth]{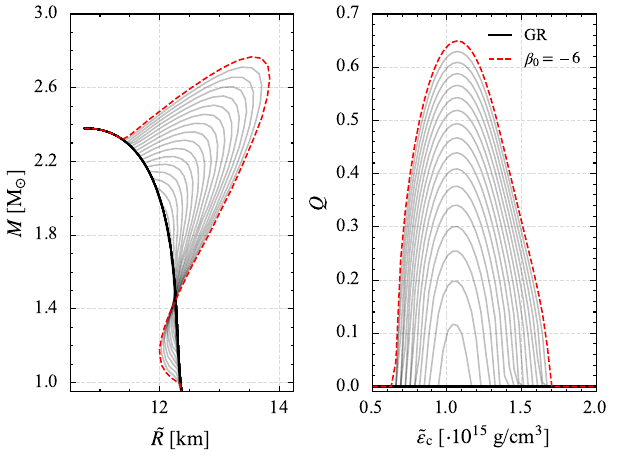}
\caption{
Neutron star scalarization. We show a sequence of neutron stars solutions
modeled using, as an example, the APR equation of state~\cite{Akmal:1998cf}, for $\alpha_0 = 10^{-5}$ and
$\beta_0 \in [-6, 0]$ in intervals of $\Delta \beta_0 = 0.1$.
The left panel shows the mass-radius relations, while the right panel shows the
charge-to-mass ratio $Q$ as a function to central density. When $\beta_0
\lesssim -4.5$, spontaneous scalarization takes place, causing the stars to
have a nonzero $Q$ and masses and radii that can deviate considerably from
general relativity.}
\label{fig:scalarization}
\end{figure}

The upshot of this exercise (see Ref.~\cite{Silva:2014fca}, Sec.~4.3., for
details) is that bound states (for $l = 0$) exist when $\beta_0 \lesssim -4.35$, quite
independently of the equation of state.
These bound states mark the bifurcation point from the general-relativistic
sequence of solutions of a new branch of neutron stars endowed with a
nontrivial scalar field.
This is known as spontaneous scalarization.\footnote{
The terminology is inspired by the phenomenon of spontaneous
magnetization of ferromagnets~\cite{Damour:1996ke}. See also
Ref.~\cite{Harada:1998ge} for an analysis of scalarization within catastrophe theory.}
Figure~\ref{fig:scalarization} shows the mass-radius relation of neutron
stars in this theory.
These curves were obtained by solving the TOV equations~\eqref{eq:tov_stt}
with the $\alpha = \beta_0 \vp$ and asymptotic boundary condition $\vp_{0} = 0$,
using a shooting method as outlined in Box~2.

The fact that the theory admits multiple branches of solutions begs the
question: which of them is favored?
To analyze this question we can look at the binding energy $E_{\rm b}$, defined in Eq.~\eqref{eq:def_binding_energy}.
One finds that stars in the scalarized branch have larger binding energy and are energetically favorable over the nonscalarized solutions~\cite{Damour:1993hw,Harada:1997mr,Horbatsch:2010hj}.
This has been confirmed in numerical simulations by Novak~\cite{Novak:1998rk},
who studied the dynamical evolution of neutron stars from the nonscalarized to
the scalarized branch.

The presence of the scalar charge causes the neutron star, when placed in a
binary, to source scalar dipole radiation. This causes the orbital motion of
the binary to evolve differently if were to take into account gravitational
radiation only. Binary pulsar observations have severely constrained the
existence of scalar dipole radiation and have essentially ruled out spontaneous
scalarization in the model proposed by Damour and
Esposito-Far\`ese~\cite{Kramer:2021jcw,Zhao:2022vig}

To close our brief discussion on scalarization, let us revisit two statements
we made here.
First, we set $\alpha_0 = 0$. If we had not done this, it is easy to see
from Eq.~\eqref{eq:eom_phi_stt} that $\vp = 0$ is no longer a solution of the
theory in the presence of matter. This means neutron stars (and any other
material object) in Brans-Dicke theory always have some scalar field.
Ultimately, this is what makes it possible to constraint this theory with Solar
System observations.
Scalarization, in contrast, requires strongly self-gravitating objects.
Second, we said that in general $\tilde{T} < 0$ and for this reason
scalarization happens when $\beta_0 < 0$. It turns out that for some equations
of state, $\tilde{T}$ can become positive at sufficiently large densities.
Hence, scalarization can also happen for positive values of $\beta_0$ in these
cases. See Refs.~\cite{Mendes:2014ufa,Mendes:2016fby,Palenzuela:2015ima,Podkowka:2018gib} for
details.

\subsubsection{The black-hole limit}
\label{sec:bh_limit}

Let us close our discussion by contrasting neutron stars with black holes in
scalar-tensor gravity.
Equations~\eqref{eq:line_element_just} and \eqref{eq:phi_just} remain valid in
a black-hole spacetime.
If Eq.~\eqref{eq:line_element_just} describes a black hole, the event
horizon $\varrho_{\hh}$ will be located at $\varrho_{\hh} = a$.
However, from Eq.~\eqref{eq:phi_just}, the scalar field $\vp$, and all its
derivatives, become singular at the horizon, unless $q = 0$.
As we saw earlier, for $q = 0$ we obtain the Schwarzschild metric plus a
trivial scalar field.

This result is nothing but a statement of the no-hair theorems in scalar-tensor
gravity.
It teaches us an important lesson: \emph{to test gravity, sometimes it is better to
rely on neutron stars rather than on black holes.}
See Ref.~\cite{Herdeiro:2015waa} for a review on no-hair theorems.

\subsection{Remarks about other gravity theories}
\label{sec:ns_others}

We saw how to construct nonrotating neutron star solutions in general
relativity and in scalar-tensor gravity.
In the latter, we found that neutron stars can be significantly different
with respect to their general-relativistic counterparts through an effect known
as spontaneous scalarization. What can happen in other gravity theories?
Before moving on, let us highlight the salient properties of neutron stars in a
few other gravity theories.

\subsubsection{Massive scalar-tensor gravity}

In this theory, one chooses the potential appearing in Eq.~\eqref{eq:action_stt} to have
the form $V(\vp) = m_{\vp}^2 \vp^2 / (2 \hbar^2)$, giving the scalar field a mass $m_{\vp}$.
This simple modification is sufficient to make the Damour-Esposito-Far\`ese model consistent
with binary pulsar observations.
The reason can be understood qualitatively as follows. The inclusion of the
mass term modifies the spatial behavior of the scalar field outside the star
causing it to decay exponentially as
$\vp \sim (1/r)\,\exp[-(m_{\vp}/\hbar)r]$,
in the far-field limit.
This strongly suppresses the scalar interaction between the binary components
when the orbital separation $R_{\rm orb}$ is larger than the Compton wavelength
\begin{equation}
\lambda_{\vp} = 2 \pi \hbar / (m_{\vp} c) \,,
\label{eq:compton}
\end{equation}
associated to the scalar field.
In this case, the emission of scalar dipole radiation is suppressed~\cite{Alsing:2011er}.
Then, to evade current binary pulsar constraints on the emission of scalar
dipole radiation one can chose $m_{\vp} \gg 10^{-16}~{\rm eV}$, and
$3 \lesssim - \beta \lesssim 10^3$
if one wants neutron stars, but not white
dwarfs, to scalarize~\cite{Ramazanoglu:2016kul}.
See also the related ``asymmetron'' theory proposed in
Refs.~\cite{Chen:2015zmx,Morisaki:2017nit}.

Regarding scalarization, the main difference with respect to
the massless case, is that the scalar field mass raises the threshold for
scalarization, for a fixed value of $\beta_0$.
Qualitatively, we can imagine that by giving the scalar field a mass we would
require more energy to excite its perturbations, hence making scalarization
harder to happen.

Nonetheless, this is compensated by the wider range of value of $\beta_0$
compatible with observations that one can work with.
A detailed investigation of neutron star solutions was done in
Ref.~\cite{Rosca-Mead:2020bzt}.
This work also discusses numerical methods to solve the TOV equations using a
relaxation scheme.

\subsubsection{Einstein-scalar-Gauss-Bonnet gravity}
\label{sec:ns_esgb}

This theory is described by the action
\begin{equation}
S = \frac{1}{16 \pi} \int \dV \left[
    R - 2 (\pd \vp)^2 + \ell^2_{\rm GB} \, f(\vp) \, R^2_{\rm GB}
    \right]
    + S_{\rm m}[\Psi_{\rm m}; g_{\mu\nu}]\,,
\label{eq:action_esgb}
\end{equation}
where $\ell_{\rm GB}$ is a lengthscale. The theory has as main feature the
coupling between the scalar field $\vp$ with the higher-order curvature scalar,
known as Gauss-Bonnet invariant:
\begin{equation}
    R^2_{\rm GB} = R^2 - 4 R_{\alpha\beta}R^{\alpha\beta} + R_{\alpha\beta\gamma\delta}R^{\alpha\beta\gamma\delta} \,.
\label{eq:gauss_bonnet}
\end{equation}
In four-dimensional spacetimes, $R_{\rm GB}^2$ is total derivative and therefore cannot contribute
to the field equations. However, the coupling of $R_{\rm GB}^2$ to $\vp$, causes the former
to contribute nontrivially to the theory's field equations.

The theory described by the action~\eqref{eq:action_esgb} has a few interesting properties
that makes it an attractive toy model to explore beyond-general-relativity physics.
Firstly, the field equations of the theory are at most second-order in time and therefore the
theory is free from negative-energy propagating degrees of freedom known as ghosts. (This is
known as the ``Ostrogradski instability''; see, e.g.,~Refs.~\cite{Kobayashi:2019hrl,Woodard:2015zca}.)
In fact, Einstein-scalar-Gauss-Bonnet gravity, is a subclass of the broader
class of Horndeski theories of gravity~\cite{Horndeski:1974wa}, the most
general scalar-tensor theory in four dimensions with this property;
see Ref.~\cite{Kobayashi:2019hrl} for a review.
Secondly, the theory introduces a single free parameter, the lengthscale $\ell_{\rm GB}$.
Thirdly, the coupling between $\vp$ and $R_{\rm GB}^2$ is present in low-energy limits of
string theory; see e.g.,~Refs.~\cite{Cano:2021rey,Metsaev:1987zx}.

The action~\eqref{eq:action_esgb} yields the following field equation for the scalar field
\begin{equation}
    \Box \vp = - \tfrac{1}{4} \ell_{\rm GB}^2 \, f_{,\vp}(\vp) \, R_{\rm GB}^2 \,,
    \label{eq:eom_phi_esgb}
\end{equation}
where $f_{,\vp}(\vp) = \dd f  / \dd \vp$.
We can identify two broad subclasses of Einstein-scalar-Gauss-Bonnet gravity
according to the form of the coupling function $f(\vp)$.
One subclass, call it Type~I, has
$f_{,\vp}(\vp) \neq 0$ for all $\vp$. Examples, include the dilatonic $f
\sim \exp(\vp)$ and the shift-symmetric $f \sim \vp$ couplings.\footnote{The name dilatonic
arises from the context of string theory, where $\vp$ is called a ``dilaton field.'' The name
``shift symmetric'' is due to the fact that for this coupling function the action is invariant under
a shift $\vp \to \vp + {\rm const.}$.}
Another subclass, say Type~II, admits $f_{,\vp}(\vp_0) = 0$ for some constant $\vp_0$. Examples include
the quadratic $f \sim \vp^2$ and Gaussian $f \sim \exp(-\vp^2)$ couplings.

Let us first briefly discuss neutron star solutions in Type I theories.
Yagi et al.~\cite{Yagi:2015oca} showed that whether or not neutron stars have monopole
scalar charge in these theories depends on the functional form of $f$.
In particular, they showed that neutron stars in the shift-symmetric
theory \emph{do not} posses monopole scalar charge, whereas in the dilatonic
case they do.
This is interesting for it identifies a subclass of theories in which scalar
dipole radiation emission would vanish in binary pulsars.
In contrast, black holes in the shift-symmetric theory \emph{do have} monopole
scalar charge~\cite{Yunes:2009ch}.
More generally, the interaction term $\vp \, R_{\rm GB}^2$ is
necessary for black holes to have monopole scalar charge in any shift-symmetric
Horndeski theory~\cite{Sotiriou:2013qea,Sotiriou:2014pfa,Saravani:2019xwx}.
This is means that to test this theory, a binary with at least one black hole
is desirable over a double neutron star system. Compare with our discussion
in Sec.~\ref{sec:bh_limit}.

We now turn to the Type II theories. Their most important feature is that they
admit spontaneous scalarization~\cite{Silva:2017uqg,Doneva:2017jop}.
We can understand why this happens by analogy with the equations we met in
Sec.~\ref{sec:neutron_star_scalarization}.
More precisely, Eq.~\eqref{eq:box_linearized_def}
can be compared against Eq.~\eqref{eq:eom_phi_esgb}, with the latter
particularized for the coupling function $f = \eta~\vp^2 / 2$ and
likewise linearized around a neutron-star background in general relativity,
\begin{equation}
    \Box \delta \vp = - \tfrac{1}{4} \eta~\ell_{\rm GB}^2 R_{\rm GB}^2 \delta \vp \,,
\end{equation}
where $\eta$ is a dimensionless constant.
We can then make the identifications
\begin{equation}
    \eta \sim 16 \pi \beta_0, \quad \textrm{and} \quad \ell_{\rm GB}^2~R_{\rm GB}^2 \sim \tilde{T} \,,
\end{equation}
establishing the analogy between the two theories.
Despite the complexity of Eq.~\eqref{eq:gauss_bonnet}, the Gauss-Bonnet invariant for
a neutron-star spacetime in general relativity is simple~\cite{Silva:2017uqg}
\begin{equation}
    R_{\rm GB}^2 = \frac{48 m^2}{r^6} - \frac{128 \pi (m + 2 \pi r^3 p) \varepsilon}{r^3} \,.
\end{equation}
We can use the near-center expansion formulas~\eqref{eq:near_core_series}, to
conclude that $R_{\rm GB}^2$ is negative at the star's core. We also see that
$R_{\rm GB}^2$ matches smoothly the Schwarzschild result
$R_{\rm GB}^2 = 48 M^2 / r^6$
at the star's surface $r = R$, where $m(r) = M$, as we should expect from the
Jebsen-Birkhoff theorem.
The fact that $R^2_{\rm GB}$ is not positive-definite suggests that
scalarization can happen for both signs of the coupling constant
$\eta$~\cite{Silva:2017uqg}.
When $\eta$ is negative, scalarization is triggered deep within the star, while
for $\eta$ positive, scalarization is triggered near the star's surface.
Scalarized neutron stars exhibit the familiar bifurcation off the
nonscalarized sequence of solutions~\cite{Doneva:2017duq}.
Stars in the scalarized branch have smaller mass (at fixed central density)
with respect to nonscalarized solution, and are energetically favored over
the latter.
Finally, the scalarized solutions have a monopole scalar charge~\cite{Doneva:2017duq}.

The most striking difference between scalarization in the Damour-Esposito-Far\`ese model
and in Type II Einstein-scalar-Gauss-Bonnet gravity is that in the latter scalarization
also occurs for black holes~\cite{Doneva:2017jop,Silva:2017uqg}. This follows from
$R^2_{\rm GB}$ being nonzero for vacuum black-hole spacetimes whereas $\tilde{T}$ is zero.
We refer the reader to Ref.~\cite{Doneva:2022ewd}, Sec.~IV, for a review of
black-hole scalarization.

\subsubsection{Dynamical Chern-Simons gravity}
\label{sec:theory_dcs}

This theory is described by the action~\cite{Jackiw:2003pm}
\begin{equation}
S = \frac{1}{16 \pi} \int \dV \left[
    R - 2 (\pd \vt)^2 + \ell^2_{\rm CS} \, \vt \, \pont
    \right]
    + S_{\rm m}[\Psi_{\rm m}; g_{\mu\nu}]\,,
\label{eq:action_dcs}
\end{equation}
where $\ell_{\rm CS}$ is a lengthscale. The theory has as main feature the
coupling between the pseudoscalar field $\vt$ with the higher-order curvature scalar,
known as the Pontryagin density
\begin{equation}
    \pont = R_{\nu\mu\rho\sigma} \, {}^{\ast}R^{\,\mu\nu\rho\sigma} \,,
\label{eq:pontryagin}
\end{equation}
where ${}^{\ast}R^{\mu\nu\rho\sigma} = \varepsilon^{\,\rho\sigma\alpha\beta}
R^{\mu\nu}{}_{\alpha\beta}/2$ is the dual of the Riemann tensor and
$\varepsilon^{\,\rho\sigma\alpha\beta}$ is the totally antisymmetric Levi-Civita tensor;
see Ref.~\cite{Alexander:2009tp} for a review.
As with the case of the Gauss-Bonnet invariant, $\pont$ is a total derivative in
four dimensions and it cannot contribute to the field equations unless we
couple it to $\vt$.
Contrary to Einstein-scalar-Gauss-Bonnet gravity, dynamical Chern-Simons theory
has field equations which are third-order in time. For this reason, the theory
should be seen as an effective theory, with $\ell_{\rm CS} / \ell \ll 1$, where $\ell$
is some characteristic lengthscale associate to the problem; see Ref.~\cite{Delsate:2014hba}.
For neutron stars, one imposes, for example, that $\ell_{\rm CS} / M \ll 1$.

An interesting aspect of this theory, is that the Pontryagin density is
zero for any spherically symmetric spacetime~\cite{Jackiw:2003pm}.
For this reason, nonrotating neutron stars, which is what we have been
considering so far, \emph{are identical to those of general relativity.}
Deviations from general relativity only appear when the neutron-star spacetime
is no longer spherically symmetric. The simplest example of such a
spacetime is when one gives a small degree of rotation to star.
In this case, the line element~\eqref{eq:line_element_sch}
is augmented by a mixed $\dd t \, \dd \phi$ term as follows:
\begin{equation}
    \dd s^2 =
    - e^{2 \nu(r)} \, \dd t^2 + e^{2 \lambda(r)} \, \dd r^2
    + r^2 \, \dd \theta^2 + r^2 \sin^2\theta \, \dd \varphi^2
    + 2 \omega(r, \theta) r^2 \sin^2\theta \, \dd t \, \dd \phi \,,
    \label{eq:line_element_spin}
\end{equation}
where $\omega$ is a function that represents the angular velocity acquired by a
particle falling from infinity as measured by a static asymptotic observer
(``frame dragging effect'')~\cite{Hartle:1967he}.
The scalar field sourced by the slowly-rotating star is not spherical
symmetric, and has instead a dipolar structure~\cite{Yunes:2009ch}.
Asymptotically, the scalar field decays as
\begin{equation}
    \vt \simeq \mu \cos\theta / r^2 + {\cal O}(r^{-3}) \,,
\end{equation}
where $\mu$ is a scalar dipole charge, whose magnitude is proportional to $\ell_{\rm
CS} / M$ and the angular momentum of the star.
This is different from the other theories we encountered so far.
See Ref.~\cite{Yagi:2013mbt} for details.

An important consequence of these results is that dynamical Chern-Simons
gravity is poorly constrained by binary pulsar observations.
This is because the binary components are slowly rotating\footnote{
Of course, pulsars are fast rotating in comparison to any other astronomical
object we know of. Here we mean ``slow'' with respect to to the mass-shedding
frequency $f_{\rm ms} = (2\pi)^{-1}~\sqrt{G M / R_{\rm eq}}$, at which the
angular velocity of the star equals that of a particle in Keplerian orbit at the star's
equator.
For a representative millisecond pulsar with $M = 1.4~\msun$, $R_{\rm eq} = 12$~km
and spinning at $f = 700$~Hz, we have $f / f_{\rm ms} \approx 0.4$.
This justifies the use of a slow-rotation expansion to model rotating neutron
stars, as pioneered by Hartle~\cite{Hartle:1967he}.
Equation~\eqref{eq:line_element_spin} is the leading order in this
expansion. See also Refs.~\cite{Hartle:1968si,Chandrasekhar:1974} and \cite{Benhar:2005gi}
for details on the expansion to second and third orders in spin, respectively.
The subject of \emph{rapidly} rotating neutron stars is presented in the monographs~\cite{Friedman:2013xza,Gourgoulhon:2010ju}.}
and the dipole scalar charge is proportional to the spin angular momentum
through the quadratic-in-curvature terms~\cite{Yagi:2013mbt}.
Nevertheless, the strongest constraint to date on this theory comes from neutron star
observations~\cite{Silva:2020acr}.
We will explain how this is possible in Sec.~\ref{sec:eos_deg}.

\section{Tests of gravity with isolated neutron stars}
\label{sec:tgr_ins}

Now that we have seen in a few examples how neutron stars are different in
theories beyond general relativity, we will proceed to discuss the
observational implications of these results. Some, such the emission of scalar
radiation in a binary, were already mentioned.
First, let us clarify what we mean by the statement of ``testing a theory of
gravity with neutron star observations''.

\subsection{Digression: what do we mean by testing gravity?}

In general, observations provides us with some data $d$ and we use a model (or
hypothesis) ${\cal H}$ to attempt to explain this data.
The model ${\cal H}$ would usually assume general relativity combined with
other physics (say hydromagnetics) that are relevant to produce an accurate
description of the observed data $d$.
Any statistical inference from $d$ using ${\cal H}$ (say, an estimate of a
neutron star mass) will come with an associated error, with a statistical and
a systematic components.
Both error components are always present. The former because our observations
are subject to noise sources and the latter because our models are not a
perfect description of nature.
Qualitatively, the statistical error gives the width of the posterior
distribution of a fiducial inferred parameter, while the systematic error tells
how much the peak of the posterior distribution (i.e., the most likely value of the parameter) disagrees from
the true value of the parameter.
We are fine as long as our systematic errors are smaller than our statistical
errors. That is because our knowledge about the value of this fiducial
parameter will still overlap with its true value (or accepted value) to some
confidence level.
However, this will no longer be the case if the systematic errors become
dominant. If this happens, our inferences become biased.

It is often the case that we make statistical inferences assuming the validity
of general relativity. We then fiddle around with the free parameters of some
alternative theory and identify how large they can be before getting in tension
with the error in our parameter inference. This gives an upper bound on the
parameters of the modified theory. Many of the results that will quote
should be interpreted in this way.

If nature happens to not be described by general relativity, this would be a
source of systematic error in any analysis that uses a model that assumes
general relativity to be true. From our previous discussion, this may lead to
a fundamental bias (using the language of Ref.~\cite{Yunes:2009ke}) in our parameter
inference. This nomenclature is used to distinguish this type of bias from other biases originating, for
instance, from some simplified or overlooked physics that went into ${\cal H}$.

In reality, of course, when we collect data with our telescopes we have no idea
whether the photons hitting the detector were produced in a process in which
general relativity or an alternative holds.
A solution to this problem is to come up with a model that is to some
extend theory agnostic.
We say ``to some extend'' because one still relies on some physical
assumptions (i.e., symmetry requirements or the hypothesis that we deviate a
little from general-relativistic predictions) to develop such model.
This is the approach adopted in the
parametrized post-Newtonian (to test gravity at first post-Newtonian order with Solar System tests)~\cite{Will:1993hxu},
the parametrized post-Keplerian~\cite{Deruelle:1986,Damour:1992} (to test gravity with binary pulsars), and
the parametrized post-Einsteinian formalism~\cite{Yunes:2009ke,Chatziioannou:2012rf} and parametrized effective-one-body waveform models~\cite{Brito:2018rfr,Ghosh:2021mrv,Maggio:2022hre} (to test gravity with gravitational waves), to name a few.
All these theory-agnostic models (and others that we will see in
Secs.~\ref{sec:eos_deg} and~\ref{sec:parametrized_tov}) have free parameters
which have specific values in general relativity. We can then include them into
our parameter inference. If their posterior distributions agree with their
predicted values in general relativity we have confirmed Einstein's theory, if not,
it signifies the presence of some systematical effect caused either by the true
value of the parameter not being the one predicted by general relativity
(exciting) or by some extra physics missed out by our formalism (often
boring.)
It often happens that these constraints on theory-agnostic parameters can be
mapped in theory-specific constraints.
One should also have in mind that sometimes missing nongravitational physics
may emulate beyond-general-relativity effects, and potentially lead to false
violations of general relativity.

Finally, one could develop also two models, ${\cal H}_{1}$ and ${\cal H}_{2}$,
which include the same physics and differing only on the theory of gravity
assumed. We will see one such example in Sec.~\ref{sec:ns_pp}. We could then
perform Bayesian model selection between the two models and see which one is
favored by the data.
Theory-agnostic and theory-specific approaches are complementary to one another
and pursuing both is a healthy approach in testing gravity.

It is important to keep in mind that sometimes scientific progress allows us to
take big leaps across different regimes of a theory.
This was certainly the case with the first detection of gravitational
waves produced by the coalescence of a black-hole binary~\cite{LIGOScientific:2016aoc}.
Until 2015, we had never explored with observations the regime of gravity in
which two strongly self-gravitating bodies move at very high speeds ($v / c
\approx 0.5$ at the late inspiral) and the gravitational field is strong and
rapidly varying in time.
A similar leap happened with the discovery of binary pulsars, which opened a
new frontier in tests of gravity and expanded our testing ground beyond the
Solar System.
When faced with these situations, it makes sense to be more cautious about the
hypothesis used in modeling our new exciting data, the theory of gravity
included.

\subsection{Difficulties in testing theories of gravity with neutron star observations}

\begin{figure}[t]
\centering
\includegraphics[width=0.75\columnwidth]{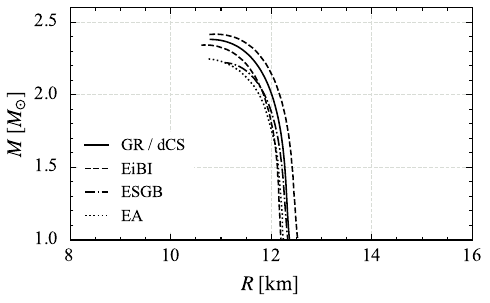}
\caption{Illustration of the gravity-theory degeneracy.
We show the mass-radius relation
for neutron stars described by the APR equation of state~\cite{Akmal:1998cf},
for a few theories of gravity, including some not covered in Sec.~\ref{sec:tgr_ins}
and values of their free constants, if any.
The theories are: general relativity (``GR,'' solid line), dynamical
Chern-Simon gravity (``dCS,'' coincident solid line with general relativity),
Eddington-inspired Born-Infeld (``EiBI,'' dashed line, $\kappa = \pm \, 5 \cdot
10^{-3}$)~\cite{Pani:2011mg}, Einstein-scalar-Gauss-Bonnet (``ESGB,'' dilatonic
coupling, dot-dashed line, $\alpha = 20~\msun$)~\cite{Pani:2011xm}, and
Einstein-Aether (``EA,'' dotted line, $c_{14} = 0.3$)~\cite{Yagi:2013ava}.
All curves terminate at the maximum-mass neutron star.
We see that even if knew the correct equation of state, it would be hard to
distinguish these theories with neutron star masses and radii
observations alone.
}
\label{fig:theory_degen}
\end{figure}

In principle, precise measurements of masses and radii of neutron stars can be
used to constrain modifications to general relativity.
Unfortunately, things are not that simple. See recent efforts in Refs.~\cite{Demirboga:2023ktt,Tuna:2022qqr}.
As we mentioned already, there are
still large uncertainties on the equation of state of neutron stars,
particularly at densities a few times larger than the nuclear saturation density
$\varepsilon_{\rm n} \simeq 2.67 \times 10^{14}$~g~cm$^{-3}$,
roughly above which the neutron-star equation of state starts becoming
poorly known.

To makes matters worse, even if we knew the correct equation of state,
different gravity theories predict similar deformations of the
general-relativistic mass-radius relation.
The only exception we are aware of are theories which allow for spontaneous
scalarization.
Figure~\ref{fig:theory_degen} illustrates this discussion.
In summary, we must face:
\begin{enumerate}
    \item \emph{The equation-of-state-gravity degeneracy:} the predictions of the bulk properties (mass, radius, moment of inertia, etc.)
    of a neutron star
    in a theory $\scrt$ belonging to the theory space $\{ \scrt_{i} \}$ are degenerate with general-relativistic predictions
    due to uncertainties in the equation of state.
    \label{itm:eos_deg}
    \item \emph{The gravity-theory degeneracy:} even if we knew the equation of state exactly,
    the prediction of the bulk properties of neutron stars for the theories in the set $\{ \scrt_{i} \}$
    are degenerate with one another.
    \label{itm:gtr_deg}
\end{enumerate}
We will discuss ways to address these two problems separately, although,
of course, both could be analyzed simultaneously.

\subsubsection{The equation-of-state-gravity degeneracy, quasiuniversal relations
and a parametrized I-Love test of gravity}
\label{sec:eos_deg}

One way to break this degeneracy is by making use of relations among neutron
star observables which depend weakly on the equation of state.
We call these ``quasiuniversal relations'' and a number of them are known to
exist in general relativity, as reviewed in Ref.~\cite{Yagi:2016bkt}.
For example, in 1989, Lattimer and Yahil~\cite{Lattimer:1989zz} proposed that
the neutron star binding energy~\eqref{eq:def_binding_energy} is given by
\begin{equation}
    E_{\rm b} \approx 1.5 \cdot 10^{51} \, (M / \msun)^2~{\rm ergs} = 0.084~\msun \, (M / \msun)^2 \,,
\label{eq:proto_eb_relation}
\end{equation}
which is valid to about 20\%.
The ``quasiuniversal'' character of Eq.~\eqref{eq:proto_eb_relation} over a
wide range of neutron star models can be improved by including the dependence on their
radii. This can be accomplished using the compactness $C$.
In particular, Lattimer and Prakash~\cite{Lattimer:2000nx} suggested the
relation
\begin{equation}
    E_{\rm b} / M = \frac{3}{5} \, \frac{C}{1 - C/2} \,.
\label{eq:eb_improved}
\end{equation}
In the same work, Lattimer and Prakash, also obtained formulas relating the
dimensionless moment of inertia $I / (MR^2)$ as a function of the compactness
$C$ by working with some exact solutions of the TOV equations in general
relativity.
For instance, they derived from what is known as the Tolman VII solution~\cite{Tolman:1939jz}
the relation,
\begin{equation}
    I / (MR^2) = \tfrac{2}{7} \, (1 - 1.1~C - 0.6~C^2 )^{-1} \,,
\end{equation}
which yields a fair approximation for $I/(MR^2)$ when $C \gtrsim 0.1$.\footnote{
The Tolman VII solution is an analytical solution of the TOV equations
that models fairly well neutron stars.
Recent works have generalized the Tolman VII solution with the inclusion of new
parameters which are then fitted against realistic equation-of-state neutron
star models, see Refs.~\cite{Jiang:2019vmf,Posada:2022lij}.
This solution was used by Ref.~\cite{Yagi:2021loe} (following
Ref.~\cite{Horbatsch:2010hj}) to derive, analytically, scalar charges in the
Damour-Esposito-Far\`ese scalarization model discussed in Sec.~\ref{sec:ns_stt}.
}
See also the earlier work of Ravenhall and Pethick~\cite{Ravenhall:1994ApJ}.
Other early quasiuniversal relations relate the oscillation and damping
timescales of pulsation modes of neutron stars to their masses and
radii~\cite{Andersson:1997rn}.
The motivation for this work is the possibility of doing neutron
star seismology. That is, infer the mass and radius of a neutron star,
and consequently learn about its internal structure, by
measuring of one or more of its pulsation modes. See
Ref.~\cite{Andersson:2021qdq} for a review.

It was later observed that the moment of inertia $I$, spin-induced quadrupole
moment $Q$, and tidal Love number $\lambda$ are remarkably equation-of-state
independent when made dimensionless by appropriate powers of the mass $M$ and
of the dimensionless spin parameter $\chi = 2 \pi f \bar{I}$, namely
\begin{equation}
\bar{I} = I / M^3, \quad \bar{Q} = - Q / (\chi^2 M^3), \quad \bar{\lambda} = \lambda / M^5.
\end{equation}
The functions $y(x)$ among any two of these three quantities are known by the
playful name ``I-Love-Q'' relations~\cite{Yagi:2013bca,Yagi:2013awa}.
For example, in its original version, $\bar{I}$ and $\bar{\lambda}$ are
related as:
\begin{equation}
\ln\bar{I} =
1.47
+ 0.0817 \ln\bar{\lambda}
+ 0.0149 \ln\bar{\lambda}^2
+ 2.87 \cdot 10^{-4} \ln\bar{\lambda}^3
- 3.64 \cdot 10^{-5} \ln\bar{\lambda}^4 \,.
\label{eq:ilq}
\end{equation}
Similar relations exist between each of these three quantities and the
compactness $C$, but to a smaller degree of equation-of-state independence.
See, e.g.,~Refs.~\cite{Breu:2016ufb,Maselli:2013mva}.

\begin{figure}[t]
\centering
\includegraphics[width=0.75\columnwidth]{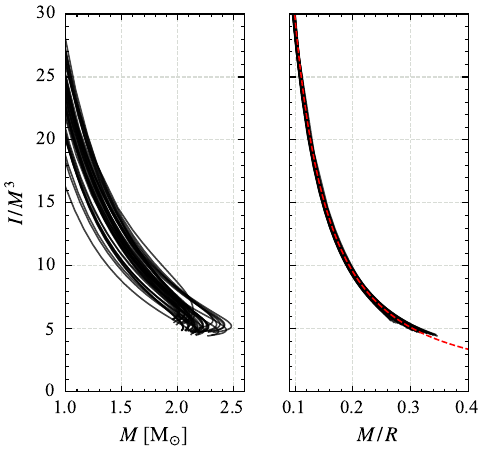}
\caption{The equation-of-state independence of the moment of inertia. In the left panel
we show the dimensionless moment of inertia $\bar{I} = I / M^3$ as a function of the mass
$M$ for several equations of state. In the right panel we show $\bar{I}$ now as a function
of the compactness $C$. Observe how the variability of $\bar{I}$ on the equation of state is
dramatically reduced when plotting it as a function of $C$. The red dashed line represents
the fit proposed in Ref.~\cite{Breu:2016ufb}, valid in the slow-rotation approximation.}
\label{fig:ic}
\end{figure}

\begin{figure}[t]
\centering
\includegraphics[width=0.75\columnwidth]{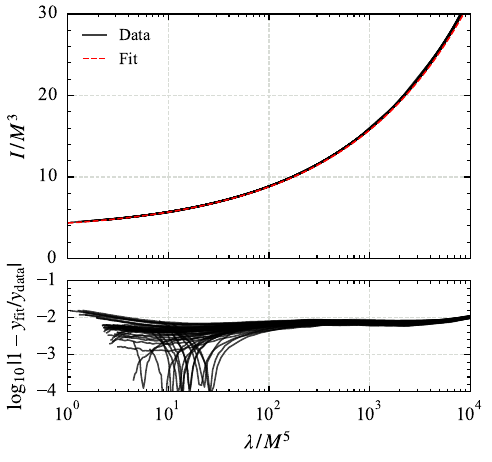}
\caption{The equation-of-state independence between dimensionless moment of inertia and
tidal Love number. In the top panel we show $\bar{I}$ and $\bar{\lambda}$ calculated for
for several equations of state (barely visible bundle of solid lines) and the fit~\eqref{eq:ilq}.
The absolute relative error between the data and fit below $\lesssim 1\%$.}
\label{fig:ilove}
\end{figure}

We illustrate our discussion in Figs.~\ref{fig:ic} and~\ref{fig:ilove}.
Figure~\ref{fig:ic} shows what we mean by equation-of-state independence.
In the left panel we show  $\bar{I}$ as a function of the mass $M$, where
each curve corresponds to a given equation of state; the same ones used in
Fig.~\ref{fig:mass_radius}.
In the right panel we plot the same curves but now as a functions of the
compactness $C$. The reduction of the variability with respect to the equation of state
is clearly visible and justifies the terminology ``quasiuniversal.''
The dashed line corresponds to the ``I-C'' relation obtained in
Ref.~\cite{Breu:2016ufb}, where the moment of inertia was calculated using the
slow-rotation expansion.
The maximum absolute relative errors, $\delta = |1 - y_{\rm fit} / y_{\rm data}|$,
between data and the fit is about $\sim 10\%$.

In Fig.~\ref{fig:ilove} we show the ``I-Love'' relation between $\bar{I}$ and $\bar{\lambda}$.
Although not apparent, the top panel contains the same number of curves as in Fig.~\ref{fig:ic},
but now the variability between different equations of state is hardly visible. The dashed line corresponds
to Eq.~\eqref{eq:ilq}. In the bottom panel we show $\delta$ between $\bar{I}_{\rm data}$ as predicted
by each equation of state with respect to the fit~\eqref{eq:ilq}.
We see that the I-Love relation holds to within $\sim 1\%$ accros all stellar models and equations of state.

An alternative to Eq.~\eqref{eq:ilq} can be worked out as follows. In Newtonian gravity, we
know that the moment of inertia $I \propto M R^2$, and hence $I/M^3 \propto C^{-2}$, while
the Love number $\lambda \propto R^5$ and hence $\lambda / M^5 \propto C^{-5}$.
Then, we can eliminate the compactness and write $\bar{I} = k \, \bar{\lambda}^{2/5}$.
Here, $k \approx 0.52$ is a constant that depends weakly on the equation of
state~\cite{Yagi:2013awa}.
This calculation can be extended systematically in powers of $C$, i.e., in a post-Minkowskian
expansion~\cite{Chan:2014tva,Jiang:2020uvb}, and yields a powers series in $C$ for both
and $\bar{I}$ and $\bar{\lambda}$. These two equations can be combined to result in
an equation for $\bar{I} (\bar{\lambda})$.
The simple relation
\begin{equation}
    \bar{I} = \bar{\lambda}^{2/5} \,
    (c_0 + c_1 \bar{\lambda}^{-1/5} + c_2 \bar{\lambda}^{-2/5} )\,,
    \label{eq:il_pm}
\end{equation}
where $c_0 = 0.584$, $c_1 = 0.980$, $c_2 = 2.965$, achieves the same degree of quasiuniversality
as Eq.~\eqref{eq:ilq}. The prefactor $\bl^{2/5}$ is the Newtonian result, and each power of
$\bl^{-1/5}$ inside the parenthesis represents a post-Minkowskian correction. The reason is that,
as we saw previously, $C \propto \bl^{-1/5}$, as long as $C$ is small~\cite{Silva:2020acr}.

Remarkably, quasiuniversal relations also hold in modified theories of gravity, see
Refs.~\cite{Doneva:2017jop} and~\cite{Yagi:2016bkt}, Sec.~6, for reviews.
However, this does not mean that they have to be described by the same formulas
as in general relativity.
This suggests that one can try to parametrize deviations from the
quasiuniversal relations in general relativity in a theory independent way.
This was done in Ref.~\cite{Silva:2020acr}, using the post-Minkowskian formula~\eqref{eq:il_pm} as baseline. The parametrized I-Love relation is
\begin{equation}
\bar{I} = \bar{\lambda}^{2/5} \,
    (c_0 + c_1 \bar{\lambda}^{-1/5} + c_2 \bar{\lambda}^{-2/5} )
    + \beta \bl^{-b/5} \,.
\label{eq:ilove_parametrized}
\end{equation}
Here, $\beta$ is a positive real number and $b$ is an integer. These parameters
encode the magnitude and the post-Minkowskian order in which the modified
gravity corrections affect the general-relativistic I-Love
relation~\eqref{eq:il_pm}, respectively.
For example, a correction at the Newtonian level corresponds to $b = -2$.

How can the I-Love relation be used to perform tests of gravity?
The first test we can perform is a null test of general relativity. Assume, for
example, that we have two independent measurements, in general relativity, of
$\bi_{1.4}$ and $\bl_{1.4}$ for a neutron with canonical mass $M = 1.4~\msun$,
and their respective errors~\cite{Yagi:2013awa}.
These measurements delimit an ``error box'' in the $\bi$-$\bl$ plane.
For the I-Love relation to be consistent with these measurements it has to pass
through this error box.
Another test we can envisage makes use of the parametrized I-Love
relation~\eqref{eq:ilove_parametrized}. For a given value of $b$, we search for
the critical value $\beta_{\rm crit}$ of $\beta$ above which the resultant
I-Love curve would be in tension with the earlier mentioned error
box~\cite{Silva:2020acr}.

\begin{figure}[t]
\centering
\includegraphics[width=0.85\columnwidth]{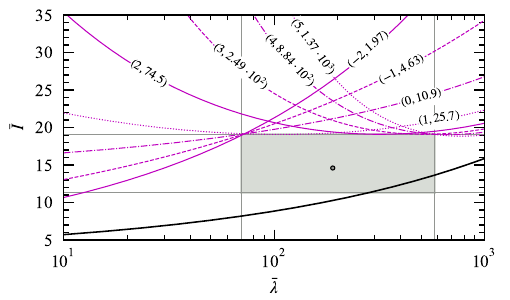}
\caption{The I-Love test of gravity. The vertical and horizontal
lines delimit the 90\% confidence region of $\bl_{1.4}$ and $\bi_{1.4}$,
while the circle marks the median $(190,\,14.6)$. The solid black line
corresponds to Eq.~\eqref{eq:il_pm} and is consistent with observation
at 90\% confidence. The various magenta curves represent the parametrized I-Love
relation~\eqref{eq:ilove_parametrized}. The labels correspond to the pair
$(b,\,\beta_{\rm crit})$. For each exponent $b$, the quantity $\beta_{\rm crit}$
represent the critical value of $\beta$ for which Eq.~\eqref{eq:ilove_parametrized}
would be in tension with the measurements of $\bi_{1.4}$ and $\bl_{1.4}$ at 90\% confidence.
Figure taken from Ref.~\cite{Silva:2020acr}.}
\label{fig:il_param}
\end{figure}

Figure~\ref{fig:il_param} shows an application of this idea. We use thevalue
inferred for the moment of inertia $\bi_{1.4} = 14.6^{+4.5}_{-3.3}$ from x-ray
observations of the pulsar J0737-3039A~\cite{Silva:2020acr}, and of the tidal
deformability $\bl_{1.4} = 190^{+390}_{-120}$ from the binary-neutron-star
gravitational-wave signal GW170817~\cite{LIGOScientific:2018cki}. The values
quoted are at 90\% confidence. The tidal deformability is obtained conditional
on the assumptions that the binary components were described by the same
equation of state and were slowly spinning.
The circle marks $(14.6,190)$ and the gray box the respective 90\% confidence
region. The solid black line is Eq.~\eqref{eq:il_pm}, which is consistent with
the two measurements. Hence, general relativity has passed this null test.
The different magenta lines correspond to post-Minkowskian exponents $b \in [-2,~5]$
and their respective values of $\beta_{\rm crit}$; both numbers are indicated in
parenthesis $(b,~\beta_{\rm crit})$ in each curve.
We remark that the constraints for $\beta \leqslant 0$ depends on the assumptions
used to infer $\bl_{1.4}$. In particular, if the tidal deformabilities were
treated as independent free parameters in the waveform model~\cite{LIGOScientific:2017vwq}, then the
$\bl_{1.4}$ posterior would not have a lower limit.
This would make all parametrized I-Love relations with $b \leqslant 0$ be consistent
with the observations.

We can wonder whether this theory-independent constraints can be translated
into constraints on any specific theory. As an example, Ref.~\cite{Silva:2020acr}
considered dynamical Chern-Simons gravity, a theory we met back in Sec.~\ref{sec:theory_dcs}.
The I-Love relation in this theory is different from that of general relativity
and comparison against a large catalog of neutron stars shows that
\begin{equation}
    b_{\rm CS} = 4\,, \quad \beta_{\rm CS} = 6.15 \times 10^{-2}\,\bar{\xi} \,,
\end{equation}
%
where $\bar{\xi} = 4 \ell_{\rm CS}^4 / M^4$, where $\ell_{\rm CS}$ is the lengthscale
associated with the theory; cf.~Eq.~\eqref{eq:action_dcs}.
For $b = 4$, we see from Fig.~\ref{fig:il_param} that $\beta_{\rm crit} = 8.84 \times 10^2$,
which can be translated into a bound on $\ell_{\rm CS}$, namely,
\begin{equation}
    \ell_{\rm CS} \leqslant 16~{\rm km} \,,
\end{equation}
at 90\% confidence, if the theory is to be consistent with the
observational bounds on $\bi_{1.4}$ and $\bl_{1.4}$.\footnote{This corresponds to $\alpha_{\rm CS}^{1/2} \leqslant 8.5$~km
in the notation used in Ref.~\cite{Silva:2020acr}. The two quantities are related by $\ell_{\rm CS} = (4\pi)^{1/4}\,\alpha_{\rm CS}^{1/2}$.}
Reference~\cite{Silva:2020acr} also showed that this value of $\ell_{\rm CS}$,
combined with the estimates on the radius and compactness of a $1.4~\msun$
neutron star derived there, satisfies the small-coupling approximation as required
by the mathematical consistency of the theory.
This is the most stringent constrain on this theory to date and is seven orders
of magnitude stronger than previous Solar System
constraints~\cite{Ali-Haimoud:2011zme,Nakamura:2018yaw}.
%
We see that quasiuniversal relations can be a powerful way to test gravity,
avoiding uncertainties regarding the equation of state.

\subsubsection{The gravity-theory degeneracy and parametrized TOV equations}
\label{sec:parametrized_tov}

The observation that different theories of gravity, given an equation of state,
yield mass-radius relations which are in general smooth deformations of the
general-relativity curve suggests that one construct a theory-agnostic
parametrized system of TOV equations.

This was the main motivation for the post-TOV formalism developed in
Refs.~\cite{Glampedakis:2015sua,Glampedakis:2016pes,Silva:2019wfn}.
The formalism is valid for static spherical stars and is inspired by the
parametrized post-Newtonian theory, extended to second PN order by
adding suitable correction terms to the fully relativistic TOV equations.
In the post-TOV formalism, the stellar structure equations are given by
\begin{subequations}
\label{eq:post_tov}
\begin{align}
    \frac{\dd p}{\dd r} &= \left( \frac{\dd p}{\dd r} \right)_{\rm GR} - \frac{\rho m}{r^2} ( \scrp_1 + \scrp_2 ) \,,
    \\
    \frac{\dd m}{\dd r} &= \left( \frac{\dd m}{\dd r} \right)_{\rm GR} + 4 \pi r^2 \rho ( \scrm_1 + \scrm_2 ) \,,
\end{align}
\end{subequations}
where $(\dd p / \dd r)_{\rm GR}$ and $(\dd m / \dd r)_{\rm GR}$ are given by Eqs.~\eqref{eq:tov_gr} and,
\begingroup
\allowdisplaybreaks
\begin{subequations}
\begin{align}
    \scrp_1 &= \delta_1 \frac{m}{r} + 4 \pi \delta_2 \frac{r^3 p}{m} \,, \\
    \scrm_1 &= \delta_3 \frac{m}{r} + \delta_4 \Pi  \,, \\
    \scrp_2 &= \pi_1 \frac{m^3}{r^5 \rho} + \pi_2 \frac{m^2}{r^2} + \pi_3 r^2 p + \pi_4 \frac{\Pi p}{\rho} \,, \\
    \scrm_2 &= \mu_1 \frac{m^3}{r^5 \rho} + \mu_2 \frac{m^2}{r^2} + \mu_3 r^2 p + \mu_4 \frac{\Pi p}{\rho} + \mu_5 \Pi^3 \frac{r}{m} \,.
\end{align}
\end{subequations}
\endgroup
In these equations, $r$ is the circumferential radius, $m$ is the mass function,
$p$ is the fluid pressure, $\rho$ is the baryonic rest mass density, $\varepsilon$
is the total energy density, and $\Pi = (\varepsilon - \rho)/\rho$
is the internal energy per unit baryonic mass.
The parameters $\delta_{i}$, $\pi_{i}$ and $\mu_{i}$ are the phenomenological post-TOV parameters.
The TOV equations are recovered when all the post-TOV parameters vanish.
The dimensionless terms $\scrp_1$, $\scrm_1$ and $\scrp_2$, $\scrm_2$ describe
the deviations from the TOV equations at 1PN and 2PN orders respectively.
In fact, the 1PN post-TOV parameters $\delta_i$ can all be expressed in terms of the more familiar
parametrized post-Newtonian parameters which are strongly constrained by observations.\footnote{Stellar solutions at 1PN order in the parametrized post-Newtonian formalism were obtained in Refs.~\cite{Ciufolini:1981,Ciufolini:1983,Shapiro:1976,WagonerMalone}. Neutron
stars were also modelled in the post-Newtonian approximation in Ref.~\cite{Shinkai:1998mg,Gupta:2000zb}, and more recently in Ref.~\cite{Andersson:2022cax}.}
In practice, one can then neglect $\scrp_1$ and $\scrm_1$ from Eqs.~\eqref{eq:post_tov}.
In contrast, $\pi_i$ and $\mu_i$ are unconstrained and $\scrp_2$ and $\scrm_2$ should be viewed
as describing the dominant departures from general relativity.

The post-TOV formalism is also in a sense ``complete'' in that, when neglecting the 1PN terms,
the post-TOV equations can be derived by the covariant conservation of a perfect fluid energy-momentum tensor.
That is,
\begin{equation}
    \nabla_{\nu} T^{\mu\nu} = 0\,, \quad T^{\mu\nu} = (\varepsilon_{\rm eff} + p)u^\mu u^\nu + p g^{\mu\nu} \,,
\end{equation}
where $\varepsilon_{\rm eff}$ is an effective, gravity-modified energy density,
\begin{equation}
    \varepsilon_{\rm eff} = \varepsilon + \rho \scrm_2 \,,
\end{equation}
and the covariant derivative is compatible with the effective post-TOV metric
\begin{equation}
    g_{\mu\nu} = \textrm{diag}[ \exp(2\nu),\, (1 - 2 m/r)^{-1},\, r^2,\, r^2 \sin^2\theta  ]\,,
\end{equation}
where the lapse $\nu$ obeys,
\begin{equation}
\frac{\dd \nu}{\dd r} = \frac{1}{r^2} \left[
    (1 - \scrm_2) \frac{m + 4 \pi r^3 p}{1 - 2m/r} + m \scrp_2
\right]\,.
\end{equation}
In the post-TOV formalism, the star's mass and radius are determined in the
same way as in general relativity.
One limitation of the post-TOV formalism is that it cannot, by construction, capture
nonperturbative effects, such as spontaneous scalarization that we saw in Sec.~\ref{sec:ns_stt}.
Some astrophysical applications of the post-TOV formalism were worked out in
Refs.~\cite{Glampedakis:2016pes,Silva:2019wfn}.

The post-TOV formalism is not the only parametrization of the TOV equations.
An ad-hoc alternative was proposed in
Ref.~\cite{Velten:2016bdk}, building on Ref.~\cite{Schwab:2008ce}.
To understand this parametrization, it is convenient to first rewrite the pressure gradient
TOV equation~\eqref{eq:tov_gr} in the equivalent factorized form
\begin{align}
    \frac{\dd p}{\dd r}
    = - (\varepsilon + p) \, \frac{\dd \nu}{\dd r}
    = - \frac{ G m \varepsilon}{r^2}
    \, \left( 1 + \vphantom{\frac{2 G m \gamma}{r}}\frac{p}{\varepsilon} \right)
    \, \left( 1 - \frac{2 G m}{r} \right)^{-1}
    \, \left( 1 + \frac{4 \pi r^3 p}{m} \right) \,.
\label{eq:dpdr_gr}
\end{align}
Each of the factors above has a specific origin. In the Newtonian limit, only the first
term survives, with the energy density $\varepsilon$ becoming  the mass density $\rho$,
i.e.,
\begin{equation}
    \frac{\dd p}{\dd r} = - \rho \frac{G m}{r^2} = - \rho \frac{\dd \Phi}{\dd r}\,,
    \label{eq:dpdr_newton}
\end{equation}
where in the second equality, $\Phi$ is the Newtonian gravitational potential,
determined from Poisson's equation.
All other terms in Eq.~\eqref{eq:tov_gr} are relativistic in origin.
The first term, $(1 + p / \varepsilon)$, acts as passive (inertial) gravitational mass
(normalized by the energy density). This can be seen by comparing the first equality in
Eq.~\eqref{eq:dpdr_gr} with the second equality in Eq.~\eqref{eq:dpdr_newton}.
The second term, $(1 + 4 \pi r^3 p / m)$, acts as the active gravitational mass density (normalized by the mass
function). This can be seen by identifying the factor $-(G m / r^2)$ in Eq.~\eqref{eq:dpdr_newton} with
$- (G m / r^2) (1 + 4 \pi r^3 p /m)$ in Eq.~\eqref{eq:dpdr_gr}.
At last, the third term, $1 / (1 - 2 G m / r)$, is a ``geometric form factor''
associated to spacetime curvature.

The idea of Refs.~\cite{Schwab:2008ce,Velten:2016bdk} is to introduce
dimensionless parameters to each of these factors. In practice, the
parametrized TOV equations are
\begin{subequations}
\label{eq:tov_velten}
\begin{align}
    \frac{\dd m}{\dd r} &= 4 \pi r^2 \varepsilon \, \left(1 + \frac{\sigma p}{\varepsilon} \right)\,,
\label{eq:dmdr_velten}
\\
\frac{\dd p}{\dd r} &=
- \frac{ G (1 + \alpha) m \varepsilon}{r^2}
\, \left( 1 - \frac{2 G m \gamma}{r} \right)^{-1}
\, \left( 1 + \vphantom{\frac{2 G m \gamma}{r}}\frac{\beta p}{\varepsilon} \right)
\, \left( 1 + \frac{4 \pi \chi r^3 p}{m} \right)
\,,
\end{align}
\end{subequations}
where $\alpha$, $\beta$, $\gamma$, $\chi$ and $\sigma$ are free constants.
The TOV equations~\eqref{eq:tov_gr} are recovered for $\alpha = \sigma = 0$ and
$\beta = \gamma = \chi = 1$. From our previous discussion, we see that $\beta$, $\gamma$ and $\chi$
affect the passive gravitational mass, geometric form factor and active gravitational
mass, respectively.
The term $\alpha$ introduces an ``effective gravitational constant'',
$G_{\rm eff} = G ( 1 + \alpha)$, whereas $\sigma$ introduces a pressure contribution
to the $\dd m / \dd r$ equation which does not exist in general relativity.

We note that neither parametrized TOV systems of equations~\eqref{eq:post_tov}
nor~\eqref{eq:tov_velten} has been generalized to rotating or tidally deformed
stars.

\subsubsection{Extreme neutron star properties}

Another way to circumvent the equation-of-state-gravity-theory degeneracy is
to look at \emph{extremal} properties of neutron stars.
The idea is simple: with few, physically sensible assumptions about the
equation of state, one uses the TOV equations to determine, for instance, what
is the maximum mass $M_{\rm max}$ a neutron star can have.

The first work in this direction was made by Rhoades and Ruffini in 1974~\cite{Rhoades:1974fn}.
Their motivation was to obtain a general upper bound on the neutron star mass,
which could then be used to distinguish them from black holes in observations
of x-ray sources in binary systems.

Specifically, Rhoades and Ruffini assumed that the star consists of two layers:
\begin{enumerate}
    \item an outer layer, whose contribution to the star's total mass $M$ is
    determined according to a known equation of state;
    \item an inner core, whose contribution to $M$ must be extremized according to a
    variational principle.
\end{enumerate}
The two layers are matched at a density $\varepsilon_0$ near the nuclear
saturation density $\varepsilon_{\rm n}$.
The variational calculation is then performed under the assumptions:
\begin{enumerate}
    \item The star is described by the TOV equations~\eqref{eq:tov_gr}.
    \label{itm:rr_ini}
    \item The energy density is positive, and related to the pressure by a barotropic equation of state~\eqref{eq:eos}.
    \label{itm:e_pos}
    \item That $\dd p / \dd \varepsilon > 0$. Thus the speed of sound $v^2_{\rm s} = \dd p / \dd \varepsilon$ is positive,
    and matter is stable against microscopic collapse.
    \item That $v_{\rm s} \leqslant 1$, and the equation of state is causal.
    \label{itm:rr_fin}
\end{enumerate}
Items~\ref{itm:e_pos} to~\ref{itm:rr_fin} are minimal physics assumptions on
the unknown equation of state at the inner core.
Rhoades and Ruffini showed that the equation of state at the core that maximizes $M$ is
\begin{equation}
    p = p_0 + (\varepsilon - \varepsilon_0), \quad \textrm{for} \quad \varepsilon \geqslant \varepsilon_0
    \label{eq:stiff_eos}
\end{equation}
where $p_0$ and $\varepsilon_0$ are the pressure and energy density at the transition density of the two stellar layers.
They found that,
\begin{equation}
    M_{\rm max} \simeq 3.2 \, \msun \,.
\end{equation}
This result depends on the value of $\varepsilon_0$ and the equation of state
assumed for the outer layer.
In the late 90s, Kalogera and Baym~\cite{Kalogera:1996ci} revisited this upper bound obtaining $M_{\rm max}
\simeq 2.9~\msun$, assuming that $\varepsilon = 2\,\varepsilon_{\rm n}$ and that
the outer layer is described by the equation of
state of Wiringa~et~al.~\cite{Wiringa:1988tp}.

Hartle and collaborators pushed this program further in the 1970's,
culminating in Ref.~\cite{Hartle:1978}.
What is particularly interesting for us, is that Ref.~\cite{Hartle:1978},
Sec.~6.3, shows a handful of extensions of this variational calculation to
other theories of gravity, most of which have now been severely constrained
by observations.\footnote{For the aficionado, the examples are: Brans-Dicke theory, Rosen's bimetric
theory and Ni's theory. The reader can find summaries about these theories in Ref.~\cite{Will:1993hxu}, Chapter 5.}

We encourage this program to be pursued further and applied to the ``modern''
gravity theories of today.
Observations of massive neutron stars consistent with general
relativity can be used to place minimally-equation-of-state-dependent
constraints on other theories.
To do so, one has to derive the prediction of $M_{\rm max}$ following
Ref.~\cite{Rhoades:1974fn} in a specific theory.\footnote{We note that
Refs.~\cite{Pani:2011xm,Sotani:2017pfj} used the equation of state~\eqref{eq:stiff_eos}
to compute the maximum neutron-star mass in modified gravity theories.
However, neither demonstrated through a calculation as in Rhoades and
Ruffini~\cite{Rhoades:1974fn} that this \emph{is} the equation of state that
maximizes the neutron star mass in the context of theories they studied.
}
Next, one looks for the regions of the theory's parameter space in which
neutron stars with $M_{\rm max} \geqslant M_{\rm obs}$ do not exist.
This would identify the regions of the theory's parameter space that would be
in tension with observations.
If we are ever to find a neutron star with a mass $M \gtrsim
3.2~\msun$, it would be hard to explain this observation within general
relativity.
It would be evidence that one (or more) of the assumptions in
Ref.~\cite{Rhoades:1974fn} is wrong; including that the TOV equations of
general relativity hold.

\section{Neutron-star pulse profiles}
\label{sec:ns_pp}

Much of our understanding of the extreme physics of neutron stars comes from
observations across the electromagnetic spectrum of astrophysical processes
taking place at the surface or at the vicinity of these stars.
See, for instance,~Refs.~\cite{Nattila:2022evn,Ozel:2012wu,Psaltis:2008bb} for reviews.
In this section, we focus on one promising way of testing theories of gravity
in the x-ray spectrum by observations of neutron-star pulse profiles, also known
as lightcurves.
These observations in the soft x-ray band made by the Neutron Star Interior Composition
Explorer (NICER) telescope are what enabled the moment of inertia inference
quoted in Sec.~\ref{sec:eos_deg}, and used in the parametrized I-Love test
described there.

The general picture is the following. By some physical process a region on the
surface of a rotating neutron stars gets heated up (relative to the rest of the
star), forming what is called a hotspot.
In general, the radiation produced by the hotspot will appear to a distant
observer as a periodic flux, modulated by the star's rotation frequency.
Photons emitted by the hotspot are subject to a number of relativistic effects
as they travel through the spacetime.
These effects, for example, gravitational time delay and gravitational
deflection leave an imprint of the properties of the neutron-star spacetime
onto the observed x-ray flux.

We saw in Sec.~\ref{sec:ns_gr} that in general relativity the exterior
spacetime is described by the Schwarzschild metric.
It is then conceivable that with sufficient sensitivity (to detect as
many photon as possible) and time-resolution (to resolve in time as well as
possible the incoming flux) one may be able to infer the mass and radius
of the radiating star.
In a gross oversimplification, this is what NICER has achieved with observations of
the isolated rotation-powered millisecond pulsar PSR J0030+0451~\cite{Miller:2019cac,Riley:2019yda} and
the massive pulsar PSR J0740+6620, the latter combining data from XMN-Newton~\cite{Miller:2021qha,Riley:2021pdl}.
The results of the former are particularly spectacular since PSR J0030+0451 is
an isolated pulsar. Two independent analysis of this system suggest
that it has mass $M = 1.44^{+0.15}_{-0.14}~\msun$ and (equatorial) radius $R = 13.02^{+1.24}_{-1.06}$~km
as inferred by~\cite{Miller:2021qha} and $M = 1.34^{+0.15}_{-0.16}~\msun$ and $R = 12.71^{+1.14}_{-1.19}$~km
as inferred by~\cite{Riley:2019yda}

\begin{figure}[t]
\centering
\includegraphics[width=0.75\columnwidth]{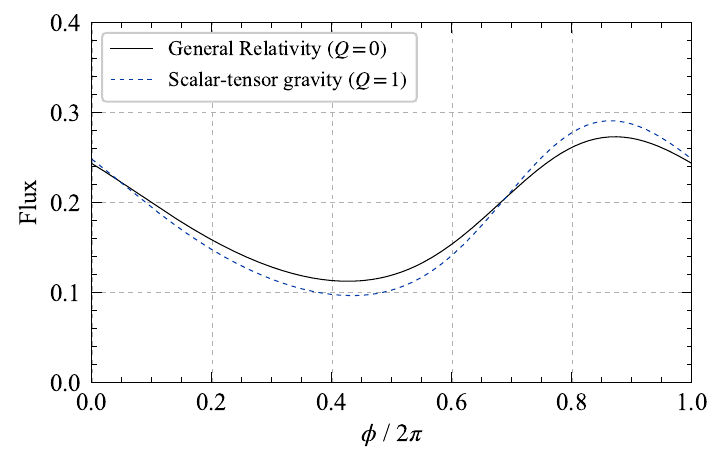}
\caption{The bolometric flux produced by a single hotpot during one revolution of
neutron star for two values of the scalar-charge-to-mass
ration $Q$ according to the J+D model.
In both cases, the star is assumed to have the same mass, $1.4~\msun$,
compactness, $C = 0.25$ and rotational frequency of $600$~Hz. All other
parameters of the model, such as the hotspot location and its temperature, are
also fixed. The only difference is the value of $Q$.
Figure taken from Ref.~\cite{Silva:2017uqg}.
}
\label{fig:pp}
\end{figure}

Given that pulse profiles are sensitive to the neutron-star spacetime, it is
natural to ask: can they be used to test general relativity and other alternatives?
To answer this question one must first develop a pulse profile model in any
given theory of gravity. In general relativity, the earliest work is this
direction dates back to 1983, when Pechenick et al.~\cite{Pechenick:1983apj}
computed the flux of circular antipodal polar caps for a slowly-rotating
neutron star including the effect of gravitational light bending, but neglecting
frame-dragging effects.\footnote{We mention in passing that the first visualization
of a neutron star surface was done by H.-P.~Nollert et al.~\cite{Nollert:1989}.
Light deflection causes the visible fraction of the star's surface to be larger
than it would be if the star was embedded in flat spacetime. This makes the
star to appear larger than it would be otherwise.
See also, e.g.,~Ref.~\cite{Nattila:2022evn}, Fig.~1, for a similar visualization
including the effects of rapid rotation obtained using the ray-tracing code of Ref.~\cite{Pihajoki:2018ihj}.}
Later works introduced special-relativistic effects of Doppler boost and
aberration caused by the rapid motion of the hotspot, and the
general-relativistic contribution due to gravitational-time delay in the
photon's arrival time at the observer.
This culminated in what is known as the ``Schwarzschild plus Doppler'' (S+D) pulse profile
model of Miller and Lamb~\cite{Miller:1997if} (see also Refs.~\cite{Poutanen:2003yd,Poutanen:2006hw}).
At high rotation frequencies ($f \gtrsim 300$~Hz), the dominant effect on the
pulse profile comes not from frame dragging, but from the spin-induced quadrupolar
deformation of the star~\cite{Braje:2000qb,Cadeau:2004gm,Cadeau:2006dc,Morsink:2007tv,Psaltis:2013zja,Nattila:2017hdb}.
One can then assume that photons are emitted from an oblate surface and then
propagate in an ambient Schwarzschild background.\footnote{The OS model also
uses an analytical description of how the star's surface deforms as a function of its mass, equatorial radius
and rotational frequency. See Refs.~\cite{AlGendy:2014eua,Morsink:2007tv,Silva:2020acr} for such formulas.}
This ``Oblate-Schwarzschild'' (OS)
approximation~\cite{Cadeau:2006dc,Morsink:2007tv} is the model used by
NICER~\cite{Bogdanov:2019qjb}.\footnote{In principle, one could do ray-tracing
on the numerical spacetime of a rotating neutron star.
In practice, due to the large parameter space of the problem, it is computationally
prohibitive to use this approach for parameter estimation using Bayesian inference.
The alternative is to use the approximate models described here.}

\begin{figure}[t]
\centering
\includegraphics[width=0.75\columnwidth]{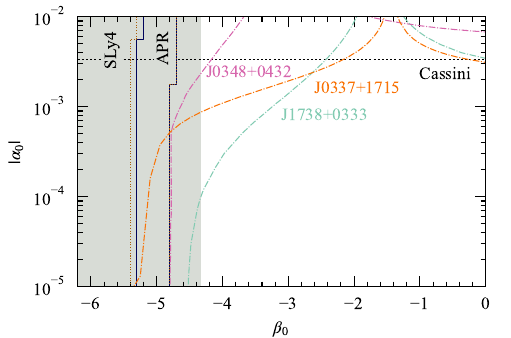}
\caption{Projected constraints on $\alpha_0$ and $\beta_0$ parameters of the scalar-tensor
gravity model of Refs.~\cite{Damour:1993hw,Damour:1996ke} using pulse profile observations.
In this plane, general relativity is located at $|\alpha_0| = 0$ and $\beta_0 = 0$.
The shaded area is the region where spontaneous scalarization happens,
and depend onle weakly on the equation of state. The region to the left of
the nearly vertical solid and dashed lines are excluded to 68\% for a star of mass $M = 1.93~\msun$,
described either by the SLy4~\cite{Douchin:2001sv} or APR~\cite{Akmal:1998cf} equation of state, spinning at 200 or 400~Hz, respectively.
For comparison, the figure also shows 68\% credible regions constrained by the Cassini
mission~\cite{Bertotti:2003rm} (dotted horizon line), the binary pulsars
J1738-0333, J0348-0432~\cite{Anderson:2019eay}, and the hierarchical triple system PSR
J0447-1715~\cite{Archibald:2018oxs}.
This figure is taken from Ref.~\cite{Silva:2019leq}. Since then, stronger
binary-pulsar constraints were obtained in Refs.~\cite{Kramer:2021jcw,Zhao:2022vig}.
These constraint strongly constrain the neutron-star scalarization, at least in the context of the original Damour-Esposito-Far\`ese model~\cite{Damour:1993hw,Damour:1996ke}.
}
\label{fig:nicer_stt}
\end{figure}

D\c{a}browski and Osarczuk~\cite{Dabrowski:1995} and Sotani and
Miyamoto~\cite{Sotani:2017bho,Sotani:2017rrt} explored how lifting the
assumption of the spacetime being Schwarzschild modifies the pulse profile.
The first model including the minimal level of astrophysical realism necessary
for pulse-profile analysis in scalar-tensor gravity was developed by Silva and
Yunes~\cite{Silva:2018yxz}. They took into account all the physical effects included in the
S+D model, assuming that the exterior spacetime is described by the Just
metric; see Sec.~\ref{sec:ns_stt}. This ``Just+Doppler'' (J+D) model has the S+D
model as the particular limit in which the scalar-charge-to-mass ratio $Q$~\eqref{eq:def_Q} becomes zero.
Figure~\ref{fig:pp} shows an example pulse profile computed within the J+D approximation.

Reference~\cite{Silva:2019leq} then used the J+D model to make a first forecast
of what constrains one could expect to place on spontaneous scalarization through
NICER-like observations consistent with general relativity.
The main conclusion of this work is that, in principle, these constraints can
be competitive with those obtained from binary-pulsar observations, as summarized in
Fig.~\ref{fig:nicer_stt}.

Despite this promising result it is fair to say that the subject has not
matured enough to give a definite answer on how useful pulse profile observations
are to test theories of gravity. For example, the analysis of Ref.~\cite{Silva:2019leq}
assumed for simplicity a single pointlike hotspot and adopted a restricted Bayesian calculation.
It would be interesting to perform a more complete analysis, similar to that of
Refs.~\cite{Lo:2013ava,Miller:2014mca}, in the J+D model to explore relative
contributions (and identify potential degeneracies) between hotspot properties
and modifications to general relativity.
Moreover, none of the beyond-general-relativity pulse profile studies
mentioned previously, as well as posterior works in massive scalar-tensor
gravity~\cite{Hu:2021tyw,Xu:2020vbs}, considered stellar
oblateness.
Eventually, of course, it would be important to use these pulse-profile models
in a Bayesian inference analysis of the data collected by
NICER~\cite{Bogdanov:2019ixe}.
It is quite clear that this subject remains a fresh, and likely fertile, ground
for work.

\section{Compact binaries and gravitational waves}
\label{sec:gws}

The first detection of gravitational waves produced by the coalescence of two
black holes on September 14, 2015, by the LIGO interferometers in Hanford and
Livingstone, USA, marked the beginning of the field of gravitational-wave
astronomy~\cite{LIGOScientific:2016aoc}.
Since then, the detectors of the LIGO-Virgo-Kagra collaboration have
detected more than 90 gravitational-wave events produced by binaries with a
black hole or a neutron star as components.
The events have been used to perform a multitude of tests of gravity, as
exemplified in Refs.~\cite{LIGOScientific:2016lio,LIGOScientific:2018dkp,LIGOScientific:2019fpa,
LIGOScientific:2020tif,LIGOScientific:2021sio,Yunes:2016jcc}.
So far, all tests have shown consistency between the predictions of general
relativity with observations.
In this section, we review the implications to tests of gravity of three
particular gravitational-wave events, where at least one binary component was
inferred to be a neutron star.

\subsection{GW170817: the first binary neutron star event}

In August 17, 2017, the LIGO and Virgo detectors observed the
gravitational-wave signal produced by the inspiral of two neutron stars. The
signal stayed approximately 100~s in the sensitivity band of the detectors and
had a signal-to-noise ratio of 32.4.
The data from the three detectors allowed a precise position in sky
localization which together with an almost simultaneous (only $\approx 1.7$~s
time delay) detection of $\gamma$-rays by the Fermi Gamma-ray Burst Monitor
enabled a massive effort to search for an electromagnetic counterpart to this
event, confirmed by several observatories.
These multimessenger observations gave support to the interpretation of
GW170817 being a neutron-star binary.

An important outcome of this event is that the almost
simultaneous observations of gravitational-wave and electromagnetic signals
that traveled a distance of approximately 130 million light years, placed the
strongest constrain to date on the propagation speed $c_g$ of gravitational
waves
\begin{equation}
    -3 \cdot 10^{-15} \leqslant c_g / c - 1 \leqslant 7 \cdot 10^{-16}.
    \label{eq:cg_bound}
\end{equation}
This result had a profound impact to extension to general relativity that are
studied as possible explanations for dark energy or as alternatives to the
standard cosmological model, $\Lambda$CDM.\footnote{This is often not the case
for the theories discussed in Sec.~\ref{sec:tgr_ins}.}
These theories generically predict that gravitational waves propagate with speed $c_g \neq c$
over a cosmological background. Hence, the strong bound~\eqref{eq:cg_bound} severely constrains
such theories.
See, e.g.,
Refs.~\cite{Baker:2017hug,Creminelli:2017sry,Ezquiaga:2017ekz,Sakstein:2017xjx}.

Let us mention two additional consequences of this event for tests of gravity.
First, the agreement between the gravitational-wave signal with the predictions
of general relativity placed a bound on the emission of scalar dipole radiation
from the system. However, this bound is weaker in comparison with those obtained with
binary pulsar observations~\cite{LIGOScientific:2018dkp,Kramer:2021jcw}.
Second, GW170817 enabled the previously mentioned estimate of the tidal
deformability of a $1.4~\msun$ neutron star, $\bar{\lambda}_{1.4} =
190^{+390}_{-120}$~\cite{LIGOScientific:2018cki} that was used to perform the
parametrized I-Love test of gravity discussed in Sec.~\ref{sec:eos_deg}.
See Ref.~\cite{Chatziioannou:2020pqz} for a review on the role of tidal
deformabality in gravitational-wave observations and its implications to
constraints on the neutron-star equation of state.

\subsection{GW200105 and GW200115: the neutron-star-black-hole binaries}

In the interval of a few days in January, 2020, the LIGO and Virgo detectors
observed the first neutron-star-black-hole binaries~\cite{LIGOScientific:2021qlt}.
These two events, GW200105 and GW200115, are interesting for tests of gravity
for the following reasons. The contribution to the gravitational-wave phase due to scalar dipole
radiation is proportional to the square of the difference of the scalar charges
of the two objects.
As we saw back in Sec.~\ref{sec:ns_esgb}, in shift-symmetric
Einstein-scalar-Gauss-Bonnet gravity neutron stars have no monopolar scalar
charge, whereas black holes do. This means that the emission of scalar dipole radiation is controlled only by the charge of the black hole, that in turn depends on $\ell_{\rm GB}$.
In addition, the relatively low mass of the black hole components in these two
events ($8.9^{+1.2}_{-1.5}~\msun$ and $5.7^{+1.8}_{-0.3}~\msun$ for GW200105 and GW200115, respectively) magnifies the effects of higher-curvature corrections, enhancing the
magnitude of the scalar charge of the black holes.
Consequently, the consistency of the data with general relativity translates
into a strong bound on this theory.
Lyu~et~al.~\cite{Lyu:2022gdr} exploited these facts to place the bound
$\ell_{\rm GB} \lesssim 5$~km using GW200115.\footnote{This corresponds to $\alpha_{\rm GB}^{1/2} \lesssim 1.33$~km
in the notation used in Ref.~\cite{Lyu:2022gdr}.
The two quantities are related by $\ell_{\rm GB}=(64 \pi)^{1/4}~\alpha_{\rm GB}^{1/2}$.}
This bound is compatible with other
astrophysical constraints on this theory~\cite{Yagi:2012gp,Saffer:2021gak} and improves the bound
$\ell_{\rm GB} \lesssim 6$~km obtained previously in Perkins et al~\cite{Perkins:2021mhb}, who
used binary black hole events that had been detected by then.
See
Refs.~\cite{Julie:2019sab,Yagi:2011xp,Shiralilou:2020gah,Shiralilou:2021mfl}
for works in post-Newtonian theory that enable these constraints to be obtained.

\section{Summary}
\label{sec:summary}

We presented an overview of how modifications to general relativity impact the
structure of neutron stars and discussed some of its astrophysical
implications.
We discussed the difficulties in testing these modified theories of gravity
using observations of neutron stars, and how these difficulties originate from
the large uncertainties in our knowledge about the neutron star equation of
state at supranucler densities. We saw how quasiuniversal relations can be
helpful to mitigate this problem, and reviewed some examples of ``theory-agnostic''
approaches to describe the structure of neutron stars.
With the hope of motivating further work, we also reviewed the current status
of pulse profile modeling in modified gravity theories, and identified some
directions for future research.
At last, we presented a small selection of constraints on modified theories of
gravity that were obtained using gravitational wave observations of compact
binaries that contain at least one neutron star.

It is not an exaggeration to say that the observations made by NICER and the
detection of gravitational-wave signals by the LIGO-Virgo-Kagra collaboration have opened a
new era in the study of neutron stars. Will these objects provide us the first
whispers of new physics beyond general relativity? This is for time to answer.
However, one thing is certain: neutron stars will continue to play a fundamental
role in our understanding of gravity at its extreme.

\begin{acknowledgement}
We thank Cosimo Bambi and Alejandro C\'ardenas-Avenda\~{n}o for the invitation
to write this chapter, and to participate in the ``Recent progress on gravity
tests'' workshop in 2022.
We are also indebted to our collaborators with whom we have studied various
aspects of tests of gravity through the neutron-star perspective over the
years.
We thank Helvi~Witek for the encouragement in the final stages of writing.
We acknowledge funding from the Deutsche Forschungsgemeinschaft (DFG) --
Project No.~386119226
\end{acknowledgement}

\bibliographystyle{myspringer}
\bibliography{biblio}

\begin{thebibliography}{100}
\providecommand{\url}[1]{\texttt{#1}}
\providecommand{\urlprefix}{URL }
\providecommand{\doi}[2][]{\href{https://doi.org/#2}{doi:#2}}
\providecommand{\eprint}[2][]{\href{https://arxiv.org/abs/#2}{arXiv:#2}}

\bibitem{LIGOScientific:2016aoc}
B.~P. Abbott \emph{et~al.}, {Observation of Gravitational Waves from a Binary
  Black Hole Merger}, Phys. Rev. Lett. \textbf{116} (2016) 061102.
\newblock \doi{10.1103/PhysRevLett.116.061102}.
\newblock \eprint{1602.03837}.

\bibitem{LIGOScientific:2016lio}
B.~P. Abbott \emph{et~al.}, {Tests of general relativity with GW150914}, Phys.
  Rev. Lett. \textbf{116} (2016) 221101.
\newblock \doi{10.1103/PhysRevLett.116.221101}.
\newblock [Erratum: Phys.Rev.Lett. 121, 129902 (2018)], \eprint{1602.03841}.

\bibitem{LIGOScientific:2017vwq}
B.~P. Abbott \emph{et~al.}, {GW170817: Observation of Gravitational Waves from
  a Binary Neutron Star Inspiral}, Phys. Rev. Lett. \textbf{119} (2017) 161101.
\newblock \doi{10.1103/PhysRevLett.119.161101}.
\newblock \eprint{1710.05832}.

\bibitem{LIGOScientific:2018cki}
B.~P. Abbott \emph{et~al.}, {GW170817: Measurements of neutron star radii and
  equation of state}, Phys. Rev. Lett. \textbf{121} (2018) 161101.
\newblock \doi{10.1103/PhysRevLett.121.161101}.
\newblock \eprint{1805.11581}.

\bibitem{LIGOScientific:2018dkp}
B.~P. Abbott \emph{et~al.}, {Tests of General Relativity with GW170817}, Phys.
  Rev. Lett. \textbf{123} (2019) 011102.
\newblock \doi{10.1103/PhysRevLett.123.011102}.
\newblock \eprint{1811.00364}.

\bibitem{LIGOScientific:2019fpa}
B.~P. Abbott \emph{et~al.}, {Tests of General Relativity with the Binary Black
  Hole Signals from the LIGO-Virgo Catalog GWTC-1}, Phys. Rev. D \textbf{100}
  (2019) 104036.
\newblock \doi{10.1103/PhysRevD.100.104036}.
\newblock \eprint{1903.04467}.

\bibitem{LIGOScientific:2021qlt}
R.~Abbott \emph{et~al.}, {Observation of Gravitational Waves from Two Neutron
  Star\textendash{}Black Hole Coalescences}, Astrophys. J. Lett. \textbf{915}
  (2021) L5.
\newblock \doi{10.3847/2041-8213/ac082e}.
\newblock \eprint{2106.15163}.

\bibitem{LIGOScientific:2020tif}
R.~Abbott \emph{et~al.}, {Tests of general relativity with binary black holes
  from the second LIGO-Virgo gravitational-wave transient catalog}, Phys. Rev.
  D \textbf{103} (2021) 122002.
\newblock \doi{10.1103/PhysRevD.103.122002}.
\newblock \eprint{2010.14529}.

\bibitem{LIGOScientific:2021sio}
R.~Abbott \emph{et~al.}
\newblock {Tests of General Relativity with GWTC-3} (2021).
\newblock \eprint{2112.06861}.

\bibitem{Akmal:1998cf}
A.~Akmal, V.~R. Pandharipande and D.~G. Ravenhall, {The Equation of state of
  nucleon matter and neutron star structure}, Phys. Rev. C \textbf{58} (1998)
  1804--1828.
\newblock \doi{10.1103/PhysRevC.58.1804}.
\newblock \eprint{nucl-th/9804027}.

\bibitem{Alexander:2009tp}
S.~Alexander and N.~Yunes, {Chern-Simons Modified General Relativity}, Phys.
  Rept. \textbf{480} (2009) 1--55.
\newblock \doi{10.1016/j.physrep.2009.07.002}.
\newblock \eprint{0907.2562}.

\bibitem{AlGendy:2014eua}
M.~AlGendy and S.~M. Morsink, {Universality of the Acceleration Due to Gravity
  on the Surface of a Rapidly Rotating Neutron Star}, Astrophys. J.
  \textbf{791} (2014) 78.
\newblock \doi{10.1088/0004-637X/791/2/78}.
\newblock \eprint{1404.0609}.

\bibitem{Ali-Haimoud:2011zme}
Y.~Ali-Haimoud and Y.~Chen, {Slowly-rotating stars and black holes in dynamical
  Chern-Simons gravity}, Phys. Rev. D \textbf{84} (2011) 124033.
\newblock \doi{10.1103/PhysRevD.84.124033}.
\newblock \eprint{1110.5329}.

\bibitem{Alsing:2011er}
J.~Alsing, E.~Berti, C.~M. Will \emph{et~al.}, {Gravitational radiation from
  compact binary systems in the massive Brans-Dicke theory of gravity}, Phys.
  Rev. D \textbf{85} (2012) 064041.
\newblock \doi{10.1103/PhysRevD.85.064041}.
\newblock \eprint{1112.4903}.

\bibitem{Alsing:2017bbc}
J.~Alsing, H.~O. Silva and E.~Berti, {Evidence for a maximum mass cut-off in
  the neutron star mass distribution and constraints on the equation of state},
  Mon. Not. Roy. Astron. Soc. \textbf{478} (2018) 1377--1391.
\newblock \doi{10.1093/mnras/sty1065}.
\newblock \eprint{1709.07889}.

\bibitem{Anderson:2019eay}
D.~Anderson, P.~Freire and N.~Yunes, {Binary pulsar constraints on massless
  scalar\textendash{}tensor theories using Bayesian statistics}, Class. Quant.
  Grav. \textbf{36} (2019) 225009.
\newblock \doi{10.1088/1361-6382/ab3a1c}.
\newblock \eprint{1901.00938}.

\bibitem{Andersson:2017kru}
N.~Andersson, {Whispers from the edge of physics}, J. Astrophys. Astron.
  \textbf{38} (2017) 58.
\newblock \doi{10.1007/s12036-017-9463-8}.
\newblock \eprint{1709.07215}.

\bibitem{Andersson:2021qdq}
N.~Andersson, {A gravitational-wave perspective on neutron-star seismology},
  Universe \textbf{7} (2021) 97.
\newblock \doi{10.3390/universe7040097}.
\newblock \eprint{2103.10223}.

\bibitem{Andersson:2022cax}
N.~Andersson, F.~Gittins, S.~Yin \emph{et~al.}, {Building post-Newtonian
  neutron stars}, Class. Quant. Grav. \textbf{40} (2023) 025016.
\newblock \doi{10.1088/1361-6382/acace5}.
\newblock \eprint{2209.05871}.

\bibitem{Andersson:1997rn}
N.~Andersson and K.~D. Kokkotas, {Towards gravitational wave asteroseismology},
  Mon. Not. Roy. Astron. Soc. \textbf{299} (1998) 1059--1068.
\newblock \doi{10.1046/j.1365-8711.1998.01840.x}.
\newblock \eprint{gr-qc/9711088}.

\bibitem{Antoniadis:2013pzd}
J.~Antoniadis \emph{et~al.}, {A Massive Pulsar in a Compact Relativistic
  Binary}, Science \textbf{340} (2013) 6131.
\newblock \doi{10.1126/science.1233232}.
\newblock \eprint{1304.6875}.

\bibitem{Archibald:2018oxs}
A.~M. Archibald, N.~V. Gusinskaia, J.~W.~T. Hessels \emph{et~al.},
  {Universality of free fall from the orbital motion of a pulsar in a stellar
  triple system}, Nature \textbf{559} (2018) 73--76.
\newblock \doi{10.1038/s41586-018-0265-1}.
\newblock \eprint{1807.02059}.

\bibitem{Baker:2017hug}
T.~Baker, E.~Bellini, P.~G. Ferreira \emph{et~al.}, {Strong constraints on
  cosmological gravity from GW170817 and GRB 170817A}, Phys. Rev. Lett.
  \textbf{119} (2017) 251301.
\newblock \doi{10.1103/PhysRevLett.119.251301}.
\newblock \eprint{1710.06394}.

\bibitem{Baym:2017whm}
G.~Baym, T.~Hatsuda, T.~Kojo \emph{et~al.}, {From hadrons to quarks in neutron
  stars: a review}, Rept. Prog. Phys. \textbf{81} (2018) 056902.
\newblock \doi{10.1088/1361-6633/aaae14}.
\newblock \eprint{1707.04966}.

\bibitem{Bedaque:2014sqa}
P.~Bedaque and A.~W. Steiner, {Sound velocity bound and neutron stars}, Phys.
  Rev. Lett. \textbf{114} (2015) 031103.
\newblock \doi{10.1103/PhysRevLett.114.031103}.
\newblock \eprint{1408.5116}.

\bibitem{Benhar:2005gi}
O.~Benhar, V.~Ferrari, L.~Gualtieri \emph{et~al.}, {Perturbative approach to
  the structure of rapidly rotating neutron stars}, Phys. Rev. D \textbf{72}
  (2005) 044028.
\newblock \doi{10.1103/PhysRevD.72.044028}.
\newblock \eprint{gr-qc/0504068}.

\bibitem{Bergmann:1968ve}
P.~G. Bergmann, {Comments on the scalar tensor theory}, Int. J. Theor. Phys.
  \textbf{1} (1968) 25--36.
\newblock \doi{10.1007/BF00668828}.

\bibitem{Berti:2015itd}
E.~Berti \emph{et~al.}, {Testing General Relativity with Present and Future
  Astrophysical Observations}, Class. Quant. Grav. \textbf{32} (2015) 243001.
\newblock \doi{10.1088/0264-9381/32/24/243001}.
\newblock \eprint{1501.07274}.

\bibitem{Bertotti:2003rm}
B.~Bertotti, L.~Iess and P.~Tortora, {A test of general relativity using radio
  links with the Cassini spacecraft}, Nature \textbf{425} (2003) 374--376.
\newblock \doi{10.1038/nature01997}.

\bibitem{Bogdanov:2019ixe}
S.~Bogdanov \emph{et~al.}, {Constraining the Neutron Star
  Mass\textendash{}Radius Relation and Dense Matter Equation of State with
  $NICER$. I. The Millisecond Pulsar X-Ray Data Set}, Astrophys. J. Lett.
  \textbf{887} (2019) L25.
\newblock \doi{10.3847/2041-8213/ab53eb}.
\newblock \eprint{1912.05706}.

\bibitem{Bogdanov:2019qjb}
S.~Bogdanov \emph{et~al.}, {Constraining the Neutron Star
  Mass\textendash{}Radius Relation and Dense Matter Equation of State with
  $NICER$. II. Emission from Hot Spots on a Rapidly Rotating Neutron Star},
  Astrophys. J. Lett. \textbf{887} (2019) L26.
\newblock \doi{10.3847/2041-8213/ab5968}.
\newblock \eprint{1912.05707}.

\bibitem{Braje:2000qb}
T.~M. Braje, R.~W. Romani and K.~P. Rauch, {Light curves of rapidly rotating
  neutron stars}, Astrophys. J. \textbf{531} (2000) 447.
\newblock \doi{10.1086/308448}.
\newblock \eprint{astro-ph/0004411}.

\bibitem{Brans:1961sx}
C.~Brans and R.~H. Dicke, {Mach's principle and a relativistic theory of
  gravitation}, Phys. Rev. \textbf{124} (1961) 925--935.
\newblock \doi{10.1103/PhysRev.124.925}.

\bibitem{Breu:2016ufb}
C.~Breu and L.~Rezzolla, {Maximum mass, moment of inertia and compactness of
  relativistic stars}, Mon. Not. Roy. Astron. Soc. \textbf{459} (2016)
  646--656.
\newblock \doi{10.1093/mnras/stw575}.
\newblock \eprint{1601.06083}.

\bibitem{Brito:2018rfr}
R.~Brito, A.~Buonanno and V.~Raymond, {Black-hole Spectroscopy by Making Full
  Use of Gravitational-Wave Modeling}, Phys. Rev. D \textbf{98} (2018) 084038.
\newblock \doi{10.1103/PhysRevD.98.084038}.
\newblock \eprint{1805.00293}.

\bibitem{Buchdahl:1959nk}
H.~A. Buchdahl, {Reciprocal Static Metrics and Scalar Fields in the General
  Theory of Relativity}, Phys. Rev. \textbf{115} (1959) 1325--1328.
\newblock \doi{10.1103/PhysRev.115.1325}.

\bibitem{Cadeau:2004gm}
C.~Cadeau, D.~A. Leahy and S.~M. Morsink, {Pulse shapes from rapidly-rotating
  neutron stars: Equatorial photon orbits}, Astrophys. J. \textbf{618} (2005)
  451--462.
\newblock \doi{10.1086/425857}.
\newblock \eprint{astro-ph/0409261}.

\bibitem{Cadeau:2006dc}
C.~Cadeau, S.~M. Morsink, D.~Leahy \emph{et~al.}, {Light Curves for
  Rapidly-Rotating Neutron Stars}, Astrophys. J. \textbf{654} (2007) 458--469.
\newblock \doi{10.1086/509103}.
\newblock \eprint{astro-ph/0609325}.

\bibitem{Cano:2021rey}
P.~A. Cano and A.~Ruip\'erez, {String gravity in D=4}, Phys. Rev. D
  \textbf{105} (2022) 044022.
\newblock \doi{10.1103/PhysRevD.105.044022}.
\newblock \eprint{2111.04750}.

\bibitem{Chan:2014tva}
T.~K. Chan, A.~P.~O. Chan and P.~T. Leung, {I-Love relations for incompressible
  stars and realistic stars}, Phys. Rev. D \textbf{91} (2015) 044017.
\newblock \doi{10.1103/PhysRevD.91.044017}.
\newblock \eprint{1411.7141}.

\bibitem{Chandrasekhar:1939isss.book}
S.~{Chandrasekhar}.
\newblock {An introduction to the study of stellar structure} (1939).

\bibitem{Chandrasekhar:1974}
S.~{Chandrasekhar} and J.~C. {Miller}, {On slowly rotating homogeneous masses
  in general relativity}, Mon. Not. Roy. Astron. Soc. \textbf{167} (1974)
  63--80.
\newblock \doi{10.1093/mnras/167.1.63}.

\bibitem{Chatziioannou:2020pqz}
K.~Chatziioannou, {Neutron star tidal deformability and equation of state
  constraints}, Gen. Rel. Grav. \textbf{52} (2020) 109.
\newblock \doi{10.1007/s10714-020-02754-3}.
\newblock \eprint{2006.03168}.

\bibitem{Chatziioannou:2012rf}
K.~Chatziioannou, N.~Yunes and N.~Cornish, {Model-Independent Test of General
  Relativity: An Extended post-Einsteinian Framework with Complete Polarization
  Content}, Phys. Rev. D \textbf{86} (2012) 022004.
\newblock \doi{10.1103/PhysRevD.86.022004}.
\newblock [Erratum: Phys.Rev.D 95, 129901 (2017)], \eprint{1204.2585}.

\bibitem{Chen:2015zmx}
P.~Chen, T.~Suyama and J.~Yokoyama, {Spontaneous scalarization: asymmetron as
  dark matter}, Phys. Rev. D \textbf{92} (2015) 124016.
\newblock \doi{10.1103/PhysRevD.92.124016}.
\newblock \eprint{1508.01384}.

\bibitem{Ciufolini:1981}
I.~{Ciufolini} and R.~{Ruffini}, {On the value of the masses of neutron stars
  in the parameterized post-Newtonian formalism}, Astron.~Astrophys.
  \textbf{97} (1981) L12--L14.

\bibitem{Ciufolini:1983}
I.~{Ciufolini} and R.~{Ruffini}, {Equilibrium configurations of neutron stars
  and the parametrized post-Newtonian metric theories of gravitation},
  Astrophys.~J. \textbf{275} (1983) 867--877.
\newblock \doi{10.1086/161580}.

\bibitem{Coquereaux:1990qs}
R.~Coquereaux and G.~Esposito-Far\`ese, {The Theory of
  Kaluza-Klein-Jordan-Thiry revisited}, Ann. Inst. H. Poincare Phys. Theor.
  \textbf{52} (1990) 113--150.

\bibitem{Creci:2023cfx}
G.~Creci, T.~Hinderer and J.~Steinhoff, {Tidal properties of neutron stars in
  scalar-tensor theories of gravity}, Phys. Rev. D \textbf{108} (2023) 124073.
\newblock \doi{10.1103/PhysRevD.108.124073}.
\newblock \eprint{2308.11323}.

\bibitem{Creminelli:2017sry}
P.~Creminelli and F.~Vernizzi, {Dark Energy after GW170817 and GRB170817A},
  Phys. Rev. Lett. \textbf{119} (2017) 251302.
\newblock \doi{10.1103/PhysRevLett.119.251302}.
\newblock \eprint{1710.05877}.

\bibitem{NANOGrav:2019jur}
H.~T. Cromartie \emph{et~al.}, {Relativistic Shapiro delay measurements of an
  extremely massive millisecond pulsar}, Nature Astron. \textbf{4} (2019)
  72--76.
\newblock \doi{10.1038/s41550-019-0880-2}.
\newblock \eprint{1904.06759}.

\bibitem{Deruelle:1986}
T.~{Damour} and N.~{Deruelle}, {General relativistic celestial mechanics of
  binary systems. II. The post-Newtonian timing formula.}, Annales de
  L'Institut Henri Poincare Section (A) Physique Theorique \textbf{44} (1986)
  263--292.

\bibitem{Damour:1992we}
T.~Damour and G.~Esposito-Far\`ese, {Tensor multiscalar theories of
  gravitation}, Class. Quant. Grav. \textbf{9} (1992) 2093--2176.
\newblock \doi{10.1088/0264-9381/9/9/015}.

\bibitem{Damour:1993hw}
T.~Damour and G.~Esposito-Far\`ese, {Nonperturbative strong field effects in
  tensor - scalar theories of gravitation}, Phys. Rev. Lett. \textbf{70} (1993)
  2220--2223.
\newblock \doi{10.1103/PhysRevLett.70.2220}.

\bibitem{Damour:1996ke}
T.~Damour and G.~Esposito-Far\`ese, {Tensor-scalar gravity and binary pulsar
  experiments}, Phys. Rev. D \textbf{54} (1996) 1474--1491.
\newblock \doi{10.1103/PhysRevD.54.1474}.
\newblock \eprint{gr-qc/9602056}.

\bibitem{Damour:1992}
T.~Damour and J.~H. Taylor, {Strong field tests of relativistic gravity and
  binary pulsars}, Phys. Rev. D \textbf{45} (1992) 1840--1868.
\newblock \doi{10.1103/PhysRevD.45.1840}.

\bibitem{Dabrowski:1995}
M.~P. {D{\c{a}}browski} and J.~{Osarczuk}, {Light curves of relativistic
  charged neutron star}, Astrophysics and Space Science \textbf{229} (1995)
  139--155.
\newblock \doi{10.1007/BF00658572}.

\bibitem{Delsate:2014hba}
T.~Delsate, D.~Hilditch and H.~Witek, {Initial value formulation of dynamical
  Chern-Simons gravity}, Phys. Rev. D \textbf{91} (2015) 024027.
\newblock \doi{10.1103/PhysRevD.91.024027}.
\newblock \eprint{1407.6727}.

\bibitem{Demirboga:2023ktt}
E.~S. Demirbo\u{g}a, Y.~E. \c{S}ahin and F.~M. Ramazano\u{g}lu, {Subtleties in
  constraining gravity theories with mass-radius data}, Phys. Rev. D
  \textbf{108} (2023) 024028.
\newblock \doi{10.1103/PhysRevD.108.024028}.
\newblock \eprint{2303.01910}.

\bibitem{Doneva:2017jop}
D.~D. Doneva and G.~Pappas, {Universal Relations and Alternative Gravity
  Theories}, Astrophys. Space Sci. Libr. \textbf{457} (2018) 737--806.
\newblock \eprint{1709.08046}.

\bibitem{Doneva:2022ewd}
D.~D. Doneva, F.~M. Ramazano\u{g}lu, H.~O. Silva \emph{et~al.}
\newblock {Scalarization} (2022).
\newblock \eprint{2211.01766}.

\bibitem{Doneva:2017duq}
D.~D. Doneva and S.~S. Yazadjiev, {Neutron star solutions with curvature
  induced scalarization in the extended Gauss-Bonnet scalar-tensor theories},
  JCAP \textbf{04} (2018) 011.
\newblock \doi{10.1088/1475-7516/2018/04/011}.
\newblock \eprint{1712.03715}.

\bibitem{Douchin:2001sv}
F.~Douchin and P.~Haensel, {A unified equation of state of dense matter and
  neutron star structure}, Astron. Astrophys. \textbf{380} (2001) 151.
\newblock \doi{10.1051/0004-6361:20011402}.
\newblock \eprint{astro-ph/0111092}.

\bibitem{Ezquiaga:2017ekz}
J.~M. Ezquiaga and M.~Zumalac\'arregui, {Dark Energy After GW170817: Dead Ends
  and the Road Ahead}, Phys. Rev. Lett. \textbf{119} (2017) 251304.
\newblock \doi{10.1103/PhysRevLett.119.251304}.
\newblock \eprint{1710.05901}.

\bibitem{Fonseca:2021wxt}
E.~Fonseca \emph{et~al.}, {Refined Mass and Geometric Measurements of the
  High-mass PSR J0740+6620}, Astrophys. J. Lett. \textbf{915} (2021) L12.
\newblock \doi{10.3847/2041-8213/ac03b8}.
\newblock \eprint{2104.00880}.

\bibitem{Franchini:2023eda}
N.~Franchini and S.~H. V\"olkel.
\newblock {Testing General Relativity with Black Hole Quasi-Normal Modes}
  (2023).
\newblock \eprint{2305.01696}.

\bibitem{Friedman:2013xza}
J.~L. Friedman and N.~Stergioulas.
\newblock {Rotating Relativistic Stars}.
\newblock Cambridge Monographs on Mathematical Physics. Cambridge University
  Press (2013).
\newblock ISBN 978-1-107-30217-4, 978-0-521-87254-6.
\newblock \doi{10.1017/CBO9780511977596}.

\bibitem{Ghosh:2021mrv}
A.~Ghosh, R.~Brito and A.~Buonanno, {Constraints on quasinormal-mode
  frequencies with LIGO-Virgo binary\textendash{}black-hole observations},
  Phys. Rev. D \textbf{103} (2021) 124041.
\newblock \doi{10.1103/PhysRevD.103.124041}.
\newblock \eprint{2104.01906}.

\bibitem{Glampedakis:2015sua}
K.~Glampedakis, G.~Pappas, H.~O. Silva \emph{et~al.},
  {Post-Tolman-Oppenheimer-Volkoff formalism for relativistic stars}, Phys.
  Rev. D \textbf{92} (2015) 024056.
\newblock \doi{10.1103/PhysRevD.92.024056}.
\newblock \eprint{1504.02455}.

\bibitem{Glampedakis:2016pes}
K.~Glampedakis, G.~Pappas, H.~O. Silva \emph{et~al.}, {Astrophysical
  applications of the post-Tolman-Oppenheimer-Volkoff formalism}, Phys. Rev. D
  \textbf{94} (2016) 044030.
\newblock \doi{10.1103/PhysRevD.94.044030}.
\newblock \eprint{1606.05106}.

\bibitem{Gourgoulhon:2007ue}
E.~Gourgoulhon.
\newblock {3+1 formalism and bases of numerical relativity} (2007).
\newblock \eprint{gr-qc/0703035}.

\bibitem{Gourgoulhon:2010ju}
E.~Gourgoulhon.
\newblock {An Introduction to the theory of rotating relativistic stars}
  (2010).
\newblock \eprint{1003.5015}.

\bibitem{Gupta:2000zb}
A.~Gupta, A.~Gopakumar, B.~R. Iyer \emph{et~al.}, {Pad\'e approximants for
  truncated postNewtonian neutron star models}, Phys. Rev. D \textbf{62} (2000)
  044038.
\newblock \doi{10.1103/PhysRevD.62.044038}.
\newblock \eprint{gr-qc/0002094}.

\bibitem{Harada:1997mr}
T.~Harada, {Stability analysis of spherically symmetric star in scalar-tensor
  theories of gravity}, Prog. Theor. Phys. \textbf{98} (1997) 359--379.
\newblock \doi{10.1143/PTP.98.359}.
\newblock \eprint{gr-qc/9706014}.

\bibitem{Harada:1998ge}
T.~Harada, {Neutron stars in scalar tensor theories of gravity and catastrophe
  theory}, Phys. Rev. D \textbf{57} (1998) 4802--4811.
\newblock \doi{10.1103/PhysRevD.57.4802}.
\newblock \eprint{gr-qc/9801049}.

\bibitem{Hartle:1967he}
J.~B. Hartle, {Slowly rotating relativistic stars. 1. Equations of structure},
  Astrophys. J. \textbf{150} (1967) 1005--1029.
\newblock \doi{10.1086/149400}.

\bibitem{Hartle:1978}
J.~B. {Hartle}, {Bounds on the mass and moment of inertia of nonrotating
  neutron stars.}, Phys.~Rept. \textbf{46} (1978) 201--247.
\newblock \doi{10.1016/0370-1573(78)90140-0}.

\bibitem{Hartle:1968si}
J.~B. Hartle and K.~S. Thorne, {Slowly Rotating Relativistic Stars. II. Models
  for Neutron Stars and Supermassive Stars}, Astrophys. J. \textbf{153} (1968)
  807.
\newblock \doi{10.1086/149707}.

\bibitem{Herdeiro:2015waa}
C.~A.~R. Herdeiro and E.~Radu, {Asymptotically flat black holes with scalar
  hair: a review}, Int. J. Mod. Phys. D \textbf{24} (2015) 1542014.
\newblock \doi{10.1142/S0218271815420146}.
\newblock \eprint{1504.08209}.

\bibitem{Hewish:1968bj}
A.~Hewish, S.~J. Bell, J.~D.~H. Pilkington \emph{et~al.}, {Observation of a
  rapidly pulsating radio source}, Nature \textbf{217} (1968) 709--713.
\newblock \doi{10.1038/217709a0}.

\bibitem{Horbatsch:2010hj}
M.~W. Horbatsch and C.~P. Burgess, {Semi-Analytic Stellar Structure in
  Scalar-Tensor Gravity}, JCAP \textbf{08} (2011) 027.
\newblock \doi{10.1088/1475-7516/2011/08/027}.
\newblock \eprint{1006.4411}.

\bibitem{Horndeski:1974wa}
G.~W. Horndeski, {Second-order scalar-tensor field equations in a
  four-dimensional space}, Int. J. Theor. Phys. \textbf{10} (1974) 363--384.
\newblock \doi{10.1007/BF01807638}.

\bibitem{Hu:2021tyw}
Z.~Hu, Y.~Gao, R.~Xu \emph{et~al.}, {Scalarized neutron stars in massive
  scalar-tensor gravity: X-ray pulsars and tidal deformability}, Phys. Rev. D
  \textbf{104} (2021) 104014.
\newblock \doi{10.1103/PhysRevD.104.104014}.
\newblock \eprint{2109.13453}.

\bibitem{Hu:2023vsq}
Z.~Hu, X.~Miao and L.~Shao.
\newblock {Tests of Classical Gravity with Radio Pulsars} (2023).
\newblock \eprint{2303.17185}.

\bibitem{Jackiw:2003pm}
R.~Jackiw and S.~Y. Pi, {Chern-Simons modification of general relativity},
  Phys. Rev. D \textbf{68} (2003) 104012.
\newblock \doi{10.1103/PhysRevD.68.104012}.
\newblock \eprint{gr-qc/0308071}.

\bibitem{Jiang:2019vmf}
N.~Jiang and K.~Yagi, {Improved Analytic Modeling of Neutron Star Interiors},
  Phys. Rev. D \textbf{99} (2019) 124029.
\newblock \doi{10.1103/PhysRevD.99.124029}.
\newblock \eprint{1904.05954}.

\bibitem{Jiang:2020uvb}
N.~Jiang and K.~Yagi, {Analytic I-Love-C relations for realistic neutron
  stars}, Phys. Rev. D \textbf{101} (2020) 124006.
\newblock \doi{10.1103/PhysRevD.101.124006}.
\newblock \eprint{2003.10498}.

\bibitem{Julie:2019sab}
F.-L. Juli\'e and E.~Berti, {Post-Newtonian dynamics and black hole
  thermodynamics in Einstein-scalar-Gauss-Bonnet gravity}, Phys. Rev. D
  \textbf{100} (2019) 104061.
\newblock \doi{10.1103/PhysRevD.100.104061}.
\newblock \eprint{1909.05258}.

\bibitem{Just:1959}
K.~{Just}, {Notizen: The Motion of Mercury according to the Theory of Thiry and
  Lichnerowicz}, Zeitschrift Naturforschung Teil A \textbf{14} (1959) 751--751.
\newblock \doi{10.1515/zna-1959-0810}.

\bibitem{Kalogera:1996ci}
V.~Kalogera and G.~Baym, {The maximum mass of a neutron star}, Astrophys. J.
  Lett. \textbf{470} (1996) L61--L64.
\newblock \doi{10.1086/310296}.
\newblock \eprint{astro-ph/9608059}.

\bibitem{Kobayashi:2019hrl}
T.~Kobayashi, {Horndeski theory and beyond: a review}, Rept. Prog. Phys.
  \textbf{82} (2019) 086901.
\newblock \doi{10.1088/1361-6633/ab2429}.
\newblock \eprint{1901.07183}.

\bibitem{Kokkotas:2000up}
K.~D. Kokkotas and J.~Ruoff, {Radial oscillations of relativistic stars},
  Astron. Astrophys. \textbf{366} (2001) 565.
\newblock \doi{10.1051/0004-6361:20000216}.
\newblock \eprint{gr-qc/0011093}.

\bibitem{Kramer:2021jcw}
M.~Kramer \emph{et~al.}, {Strong-Field Gravity Tests with the Double Pulsar},
  Phys. Rev. X \textbf{11} (2021) 041050.
\newblock \doi{10.1103/PhysRevX.11.041050}.
\newblock \eprint{2112.06795}.

\bibitem{Lattimer:2021emm}
J.~M. Lattimer, {Neutron Stars and the Nuclear Matter Equation of State}, Ann.
  Rev. Nucl. Part. Sci. \textbf{71} (2021) 433--464.
\newblock \doi{10.1146/annurev-nucl-102419-124827}.

\bibitem{Lattimer:2000nx}
J.~M. Lattimer and M.~Prakash, {Neutron star structure and the equation of
  state}, Astrophys. J. \textbf{550} (2001) 426.
\newblock \doi{10.1086/319702}.
\newblock \eprint{astro-ph/0002232}.

\bibitem{Lattimer:1989zz}
J.~M. Lattimer and A.~Yahil, {Analysis of the neutrino events from supernova
  1987A}, Astrophys. J. \textbf{340} (1989) 426--434.
\newblock \doi{10.1086/167404}.

\bibitem{Lo:2013ava}
K.~H. Lo, M.~Coleman~Miller, S.~Bhattacharyya \emph{et~al.}, {Determining
  neutron star masses and radii using energy-resolved waveforms of X-ray burst
  oscillations}, Astrophys. J. \textbf{776} (2013) 19.
\newblock \doi{10.1088/0004-637X/776/1/19}.
\newblock [Erratum: Astrophys.J. 854, 187 (2018)], \eprint{1801.08031}.

\bibitem{Lyu:2022gdr}
Z.~Lyu, N.~Jiang and K.~Yagi, {Constraints on Einstein-dilation-Gauss-Bonnet
  gravity from black hole-neutron star gravitational wave events}, Phys. Rev. D
  \textbf{105} (2022) 064001.
\newblock \doi{10.1103/PhysRevD.105.064001}.
\newblock [Erratum: Phys.Rev.D 106, 069901 (2022), Erratum: Phys.Rev.D 106,
  069901 (2022)], \eprint{2201.02543}.

\bibitem{Maggio:2022hre}
E.~Maggio, H.~O. Silva, A.~Buonanno \emph{et~al.}, {Tests of general relativity
  in the nonlinear regime: A parametrized plunge-merger-ringdown gravitational
  waveform model}, Phys. Rev. D \textbf{108} (2023) 024043.
\newblock \doi{10.1103/PhysRevD.108.024043}.
\newblock \eprint{2212.09655}.

\bibitem{Maselli:2013mva}
A.~Maselli, V.~Cardoso, V.~Ferrari \emph{et~al.},
  {Equation-of-state-independent relations in neutron stars}, Phys. Rev. D
  \textbf{88} (2013) 023007.
\newblock \doi{10.1103/PhysRevD.88.023007}.
\newblock \eprint{1304.2052}.

\bibitem{Mendes:2014ufa}
R.~F.~P. Mendes, {Possibility of setting a new constraint to scalar-tensor
  theories}, Phys. Rev. D \textbf{91} (2015) 064024.
\newblock \doi{10.1103/PhysRevD.91.064024}.
\newblock \eprint{1412.6789}.

\bibitem{Mendes:2016fby}
R.~F.~P. Mendes and N.~Ortiz, {Highly compact neutron stars in scalar-tensor
  theories of gravity: Spontaneous scalarization versus gravitational
  collapse}, Phys. Rev. D \textbf{93} (2016) 124035.
\newblock \doi{10.1103/PhysRevD.93.124035}.
\newblock \eprint{1604.04175}.

\bibitem{Metsaev:1987zx}
R.~R. Metsaev and A.~A. Tseytlin, {Order alpha-prime (Two Loop) Equivalence of
  the String Equations of Motion and the Sigma Model Weyl Invariance
  Conditions: Dependence on the Dilaton and the Antisymmetric Tensor}, Nucl.
  Phys. B \textbf{293} (1987) 385--419.
\newblock \doi{10.1016/0550-3213(87)90077-0}.

\bibitem{Miller:1997if}
M.~C. Miller and F.~K. Lamb, {Bounds on the compactness of neutron stars from
  brightness oscillations during x-ray bursts}, Astrophys. J. Lett.
  \textbf{499} (1998) L37.
\newblock \doi{10.1086/311335}.
\newblock \eprint{astro-ph/9711325}.

\bibitem{Miller:2014mca}
M.~C. Miller and F.~K. Lamb, {Determining Neutron Star Properties by Fitting
  Oblate-star Waveform Models to X-ray Burst Oscillations}, Astrophys. J.
  \textbf{808} (2015) 31.
\newblock \doi{10.1088/0004-637X/808/1/31}.
\newblock \eprint{1407.2579}.

\bibitem{Miller:2019cac}
M.~C. Miller \emph{et~al.}, {PSR J0030+0451 Mass and Radius from $NICER$ Data
  and Implications for the Properties of Neutron Star Matter}, Astrophys. J.
  Lett. \textbf{887} (2019) L24.
\newblock \doi{10.3847/2041-8213/ab50c5}.
\newblock \eprint{1912.05705}.

\bibitem{Miller:2021qha}
M.~C. Miller \emph{et~al.}, {The Radius of PSR J0740+6620 from NICER and
  XMM-Newton Data}, Astrophys. J. Lett. \textbf{918} (2021) L28.
\newblock \doi{10.3847/2041-8213/ac089b}.
\newblock \eprint{2105.06979}.

\bibitem{Morisaki:2017nit}
S.~Morisaki and T.~Suyama, {Spontaneous scalarization with an extremely massive
  field and heavy neutron stars}, Phys. Rev. D \textbf{96} (2017) 084026.
\newblock \doi{10.1103/PhysRevD.96.084026}.
\newblock \eprint{1707.02809}.

\bibitem{Morsink:2007tv}
S.~M. Morsink, D.~A. Leahy, C.~Cadeau \emph{et~al.}, {The Oblate Schwarzschild
  Approximation for Light Curves of Rapidly Rotating Neutron Stars}, Astrophys.
  J. \textbf{663} (2007) 1244--1251.
\newblock \doi{10.1086/518648}.
\newblock \eprint{astro-ph/0703123}.

\bibitem{Nakamura:2018yaw}
Y.~Nakamura, D.~Kikuchi, K.~Yamada \emph{et~al.}, {Weakly-gravitating objects
  in dynamical Chern\textendash{}Simons gravity and constraints with gravity
  probe B}, Class. Quant. Grav. \textbf{36} (2019) 105006.
\newblock \doi{10.1088/1361-6382/ab04c5}.
\newblock \eprint{1810.13313}.

\bibitem{Nattila:2022evn}
J.~N\"attil\"a and J.~J.~E. Kajava.
\newblock {Fundamental physics with neutron stars} (2022).
\newblock \eprint{2211.15721}.

\bibitem{Nattila:2017hdb}
J.~N\"attil\"a and P.~Pihajoki, {Radiation from rapidly rotating oblate neutron
  stars}, Astron. Astrophys. \textbf{615} (2018) A50.
\newblock \doi{10.1051/0004-6361/201630261}.
\newblock \eprint{1709.07292}.

\bibitem{Nollert:1989}
H.~P. {Nollert}, H.~{Ruder}, H.~{Herold} \emph{et~al.}, {The relativistic
  'looks' of a neutron star}, Astron.~Astrophys. \textbf{208} (1989) 153--156.

\bibitem{Novak:1998rk}
J.~Novak, {Neutron star transition to strong scalar field state in tensor
  scalar gravity}, Phys. Rev. D \textbf{58} (1998) 064019.
\newblock \doi{10.1103/PhysRevD.58.064019}.
\newblock \eprint{gr-qc/9806022}.

\bibitem{Oertel:2016bki}
M.~Oertel, M.~Hempel, T.~Kl\"ahn \emph{et~al.}, {Equations of state for
  supernovae and compact stars}, Rev. Mod. Phys. \textbf{89} (2017) 015007.
\newblock \doi{10.1103/RevModPhys.89.015007}.
\newblock \eprint{1610.03361}.

\bibitem{Olmo:2019flu}
G.~J. Olmo, D.~Rubiera-Garcia and A.~Wojnar, {Stellar structure models in
  modified theories of gravity: Lessons and challenges}, Phys. Rept.
  \textbf{876} (2020) 1--75.
\newblock \doi{10.1016/j.physrep.2020.07.001}.
\newblock \eprint{1912.05202}.

\bibitem{Oppenheimer:1939ne}
J.~R. Oppenheimer and G.~M. Volkoff, {On massive neutron cores}, Phys. Rev.
  \textbf{55} (1939) 374--381.
\newblock \doi{10.1103/PhysRev.55.374}.

\bibitem{Ozel:2012wu}
F.~Ozel, {Surface Emission from Neutron Stars and Implications for the Physics
  of their Interiors}, Rept. Prog. Phys. \textbf{76} (2013) 016901.
\newblock \doi{10.1088/0034-4885/76/1/016901}.
\newblock \eprint{1210.0916}.

\bibitem{Palenzuela:2015ima}
C.~Palenzuela and S.~L. Liebling, {Constraining scalar-tensor theories of
  gravity from the most massive neutron stars}, Phys. Rev. D \textbf{93} (2016)
  044009.
\newblock \doi{10.1103/PhysRevD.93.044009}.
\newblock \eprint{1510.03471}.

\bibitem{Pani:2014jra}
P.~Pani and E.~Berti, {Slowly rotating neutron stars in scalar-tensor
  theories}, Phys. Rev. D \textbf{90} (2014) 024025.
\newblock \doi{10.1103/PhysRevD.90.024025}.
\newblock \eprint{1405.4547}.

\bibitem{Pani:2011xm}
P.~Pani, E.~Berti, V.~Cardoso \emph{et~al.}, {Compact stars in alternative
  theories of gravity. Einstein-Dilaton-Gauss-Bonnet gravity}, Phys. Rev. D
  \textbf{84} (2011) 104035.
\newblock \doi{10.1103/PhysRevD.84.104035}.
\newblock \eprint{1109.0928}.

\bibitem{Pani:2011mg}
P.~Pani, V.~Cardoso and T.~Delsate, {Compact stars in Eddington inspired
  gravity}, Phys. Rev. Lett. \textbf{107} (2011) 031101.
\newblock \doi{10.1103/PhysRevLett.107.031101}.
\newblock \eprint{1106.3569}.

\bibitem{Pechenick:1983apj}
K.~R. {Pechenick}, C.~{Ftaclas} and J.~M. {Cohen}, {Hot spots on neutron stars
  - The near-field gravitational lens}, Astrophys.~J. \textbf{274} (1983)
  846--857.
\newblock \doi{10.1086/161498}.

\bibitem{Perkins:2021mhb}
S.~E. Perkins, R.~Nair, H.~O. Silva \emph{et~al.}, {Improved gravitational-wave
  constraints on higher-order curvature theories of gravity}, Phys. Rev. D
  \textbf{104} (2021) 024060.
\newblock \doi{10.1103/PhysRevD.104.024060}.
\newblock \eprint{2104.11189}.

\bibitem{Pihajoki:2018ihj}
P.~Pihajoki, M.~Mannerkoski, J.~N\"attil\"a \emph{et~al.}, {General purpose
  ray-tracing and polarized radiative transfer in General Relativity},
  Astrophys. J. \textbf{863} (2018) 8.
\newblock \doi{10.3847/1538-4357/aacea0}.
\newblock \eprint{1804.04670}.

\bibitem{Podkowka:2018gib}
D.~M. Podkowka, R.~F.~P. Mendes and E.~Poisson, {Trace of the energy-momentum
  tensor and macroscopic properties of neutron stars}, Phys. Rev. D \textbf{98}
  (2018) 064057.
\newblock \doi{10.1103/PhysRevD.98.064057}.
\newblock \eprint{1807.01565}.

\bibitem{Posada:2022lij}
C.~Posada, J.~Hlad\'\i{}k and Z.~Stuchl\'\i{}k, {New interior model of neutron
  stars}, Phys. Rev. D \textbf{105} (2022) 104020.
\newblock \doi{10.1103/PhysRevD.105.104020}.
\newblock \eprint{2201.05209}.

\bibitem{Poutanen:2006hw}
J.~Poutanen and A.~M. Beloborodov, {Pulse profiles of millisecond pulsars and
  their Fourier amplitudes}, Mon. Not. Roy. Astron. Soc. \textbf{373} (2006)
  836--844.
\newblock \doi{10.1111/j.1365-2966.2006.11088.x}.
\newblock \eprint{astro-ph/0608663}.

\bibitem{Poutanen:2003yd}
J.~Poutanen and M.~Gierlinski, {On the nature of the x-ray emission from the
  accreting millisecond pulsar SAX J1808.4-3658}, Mon. Not. Roy. Astron. Soc.
  \textbf{343} (2003) 1301.
\newblock \doi{10.1046/j.1365-8711.2003.06773.x}.
\newblock \eprint{astro-ph/0303084}.

\bibitem{Psaltis:2008bb}
D.~Psaltis, {Probes and Tests of Strong-Field Gravity with Observations in the
  Electromagnetic Spectrum}, Living Rev. Rel. \textbf{11} (2008) 9.
\newblock \doi{10.12942/lrr-2008-9}.
\newblock \eprint{0806.1531}.

\bibitem{Psaltis:2013zja}
D.~Psaltis and F.~\"Ozel, {Pulse Profiles from Spinning Neutron Stars in the
  Hartle-Thorne Approximation}, Astrophys. J. \textbf{792} (2014) 87.
\newblock \doi{10.1088/0004-637X/792/2/87}.
\newblock \eprint{1305.6615}.

\bibitem{Ramazanoglu:2016kul}
F.~M. Ramazano\u{g}lu and F.~Pretorius, {Spontaneous Scalarization with Massive
  Fields}, Phys. Rev. D \textbf{93} (2016) 064005.
\newblock \doi{10.1103/PhysRevD.93.064005}.
\newblock \eprint{1601.07475}.

\bibitem{Rangamani:2009xk}
M.~Rangamani, {Gravity and Hydrodynamics: Lectures on the fluid-gravity
  correspondence}, Class. Quant. Grav. \textbf{26} (2009) 224003.
\newblock \doi{10.1088/0264-9381/26/22/224003}.
\newblock \eprint{0905.4352}.

\bibitem{Ravenhall:1994ApJ}
D.~G. {Ravenhall} and C.~J. {Pethick}, {Neutron Star Moments of Inertia},
  Astrophys.~J. \textbf{424} (1994) 846.
\newblock \doi{10.1086/173935}.

\bibitem{Rhoades:1974fn}
C.~E. Rhoades, Jr. and R.~Ruffini, {Maximum mass of a neutron star}, Phys. Rev.
  Lett. \textbf{32} (1974) 324--327.
\newblock \doi{10.1103/PhysRevLett.32.324}.

\bibitem{Riley:2019yda}
T.~E. Riley \emph{et~al.}, {A $NICER$ View of PSR J0030+0451: Millisecond
  Pulsar Parameter Estimation}, Astrophys. J. Lett. \textbf{887} (2019) L21.
\newblock \doi{10.3847/2041-8213/ab481c}.
\newblock \eprint{1912.05702}.

\bibitem{Riley:2021pdl}
T.~E. Riley \emph{et~al.}, {A NICER View of the Massive Pulsar PSR J0740+6620
  Informed by Radio Timing and XMM-Newton Spectroscopy}, Astrophys. J. Lett.
  \textbf{918} (2021) L27.
\newblock \doi{10.3847/2041-8213/ac0a81}.
\newblock \eprint{2105.06980}.

\bibitem{Rosca-Mead:2020bzt}
R.~Rosca-Mead, C.~J. Moore, U.~Sperhake \emph{et~al.}, {Structure of neutron
  stars in massive scalar-tensor gravity}, Symmetry \textbf{12} (2020) 1384.
\newblock \doi{10.3390/sym12091384}.
\newblock \eprint{2007.14429}.

\bibitem{Saffer:2021gak}
A.~Saffer and K.~Yagi, {Tidal deformabilities of neutron stars in
  scalar-Gauss-Bonnet gravity and their applications to multimessenger tests of
  gravity}, Phys. Rev. D \textbf{104} (2021) 124052.
\newblock \doi{10.1103/PhysRevD.104.124052}.
\newblock \eprint{2110.02997}.

\bibitem{Sakstein:2017xjx}
J.~Sakstein and B.~Jain, {Implications of the Neutron Star Merger GW170817 for
  Cosmological Scalar-Tensor Theories}, Phys. Rev. Lett. \textbf{119} (2017)
  251303.
\newblock \doi{10.1103/PhysRevLett.119.251303}.
\newblock \eprint{1710.05893}.

\bibitem{Salmona:1967zz}
A.~Salmona, {Effect of Gravitational Scalar Field on High-Density Star
  Structure}, Phys. Rev. \textbf{154} (1967) 1218--1223.
\newblock \doi{10.1103/PhysRev.154.1218}.

\bibitem{Saravani:2019xwx}
M.~Saravani and T.~P. Sotiriou, {Classification of shift-symmetric Horndeski
  theories and hairy black holes}, Phys. Rev. D \textbf{99} (2019) 124004.
\newblock \doi{10.1103/PhysRevD.99.124004}.
\newblock \eprint{1903.02055}.

\bibitem{Schwab:2008ce}
J.~Schwab, S.~A. Hughes and S.~Rappaport.
\newblock {The Self-Gravity of Pressure in Neutron Stars} (2008).
\newblock \eprint{0806.0798}.

\bibitem{Shapiro:1976}
S.~T. {Shapiro} and A.~P. {Lightman}, {Rapidly rotating, post-Newtonian neutron
  stars.}, Astrophys.~J. \textbf{207} (1976) 263--278.
\newblock \doi{10.1086/154490}.

\bibitem{Shinkai:1998mg}
H.-a. Shinkai, {Truncated postNewtonian neutron star model}, Phys. Rev. D
  \textbf{60} (1999) 067504.
\newblock \doi{10.1103/PhysRevD.60.067504}.
\newblock \eprint{gr-qc/9807008}.

\bibitem{Shiralilou:2020gah}
B.~Shiralilou, T.~Hinderer, S.~Nissanke \emph{et~al.}, {Nonlinear curvature
  effects in gravitational waves from inspiralling black hole binaries}, Phys.
  Rev. D \textbf{103} (2021) L121503.
\newblock \doi{10.1103/PhysRevD.103.L121503}.
\newblock \eprint{2012.09162}.

\bibitem{Shiralilou:2021mfl}
B.~Shiralilou, T.~Hinderer, S.~M. Nissanke \emph{et~al.}, {Post-Newtonian
  gravitational and scalar waves in scalar-Gauss\textendash{}Bonnet gravity},
  Class. Quant. Grav. \textbf{39} (2022) 035002.
\newblock \doi{10.1088/1361-6382/ac4196}.
\newblock \eprint{2105.13972}.

\bibitem{Shklovsky:1967}
I.~S. {Shklovsky}, {On the Nature of the Source of X-Ray Emission of Sco
  XR-1.}, Astrophys.~J.~Lett. \textbf{148} (1967) L1.
\newblock \doi{10.1086/180001}.

\bibitem{Silva:2020acr}
H.~O. Silva, A.~M. Holgado, A.~C\'ardenas-Avenda\~no \emph{et~al.},
  {Astrophysical and theoretical physics implications from multimessenger
  neutron star observations}, Phys. Rev. Lett. \textbf{126} (2021) 181101.
\newblock \doi{10.1103/PhysRevLett.126.181101}.
\newblock \eprint{2004.01253}.

\bibitem{Silva:2014fca}
H.~O. Silva, C.~F.~B. Macedo, E.~Berti \emph{et~al.}, {Slowly rotating
  anisotropic neutron stars in general relativity and scalar\textendash{}tensor
  theory}, Class. Quant. Grav. \textbf{32} (2015) 145008.
\newblock \doi{10.1088/0264-9381/32/14/145008}.
\newblock \eprint{1411.6286}.

\bibitem{Silva:2017uqg}
H.~O. Silva, J.~Sakstein, L.~Gualtieri \emph{et~al.}, {Spontaneous
  scalarization of black holes and compact stars from a Gauss-Bonnet coupling},
  Phys. Rev. Lett. \textbf{120} (2018) 131104.
\newblock \doi{10.1103/PhysRevLett.120.131104}.
\newblock \eprint{1711.02080}.

\bibitem{Silva:2019wfn}
H.~O. Silva and N.~Yunes, {More than the sum of its parts: Combining
  parametrized tests of extreme gravity}, Phys. Rev. D \textbf{100} (2019)
  084034.
\newblock \doi{10.1103/PhysRevD.100.084034}.
\newblock \eprint{1906.00485}.

\bibitem{Silva:2019leq}
H.~O. Silva and N.~Yunes, {Neutron star pulse profile observations as extreme
  gravity probes}, Class. Quant. Grav. \textbf{36} (2019) 17LT01.
\newblock \doi{10.1088/1361-6382/ab3560}.
\newblock \eprint{1902.10269}.

\bibitem{Silva:2018yxz}
H.~O. Silva and N.~Yunes, {Neutron star pulse profiles in scalar-tensor
  theories of gravity}, Phys. Rev. D \textbf{99} (2019) 044034.
\newblock \doi{10.1103/PhysRevD.99.044034}.
\newblock \eprint{1808.04391}.

\bibitem{Sotani:2017rrt}
H.~Sotani, {Pulse profiles from a pulsar in scalar-tensor gravity}, Phys. Rev.
  D \textbf{96} (2017) 104010.
\newblock \doi{10.1103/PhysRevD.96.104010}.
\newblock \eprint{1710.10596}.

\bibitem{Sotani:2017pfj}
H.~Sotani and K.~D. Kokkotas, {Maximum mass limit of neutron stars in
  scalar-tensor gravity}, Phys. Rev. D \textbf{95} (2017) 044032.
\newblock \doi{10.1103/PhysRevD.95.044032}.
\newblock \eprint{1702.00874}.

\bibitem{Sotani:2017bho}
H.~Sotani and U.~Miyamoto, {Sensitivity of pulsar light curves to spacetime
  geometry and efficacy of analytic approximations}, Phys. Rev. D \textbf{96}
  (2017) 104018.
\newblock \doi{10.1103/PhysRevD.96.104018}.
\newblock \eprint{1710.08581}.

\bibitem{Sotiriou:2013qea}
T.~P. Sotiriou and S.-Y. Zhou, {Black hole hair in generalized scalar-tensor
  gravity}, Phys. Rev. Lett. \textbf{112} (2014) 251102.
\newblock \doi{10.1103/PhysRevLett.112.251102}.
\newblock \eprint{1312.3622}.

\bibitem{Sotiriou:2014pfa}
T.~P. Sotiriou and S.-Y. Zhou, {Black hole hair in generalized scalar-tensor
  gravity: An explicit example}, Phys. Rev. D \textbf{90} (2014) 124063.
\newblock \doi{10.1103/PhysRevD.90.124063}.
\newblock \eprint{1408.1698}.

\bibitem{Tolman:1939jz}
R.~C. Tolman, {Static solutions of Einstein's field equations for spheres of
  fluid}, Phys. Rev. \textbf{55} (1939) 364--373.
\newblock \doi{10.1103/PhysRev.55.364}.

\bibitem{Tooper:1964}
R.~F. {Tooper}, {General Relativistic Polytropic Fluid Spheres.}, Astrophysical
  Journal \textbf{140} (1964) 434.
\newblock \doi{10.1086/147939}.

\bibitem{Tooper:1966}
R.~F. {Tooper}, {Adiabatic Fluid Spheres in General Relativity.}, Astrophysical
  Journal \textbf{142} (1965) 1541.
\newblock \doi{10.1086/148435}.

\bibitem{Tuna:2022qqr}
S.~Tuna, K.~I. \"Unl\"ut\"urk and F.~M. Ramazano\u{g}lu, {Constraining
  scalar-tensor theories using neutron star mass and radius measurements},
  Phys. Rev. D \textbf{105} (2022) 124070.
\newblock \doi{10.1103/PhysRevD.105.124070}.
\newblock \eprint{2204.02138}.

\bibitem{Typel:2013rza}
S.~Typel, M.~Oertel and T.~Kl\"ahn, {CompOSE CompStar online supernova
  equations of state harmonising the concert of nuclear physics and
  astrophysics compose.obspm.fr}, Phys. Part. Nucl. \textbf{46} (2015)
  633--664.
\newblock \doi{10.1134/S1063779615040061}.
\newblock \eprint{1307.5715}.

\bibitem{CompOSECoreTeam:2022ddl}
S.~Typel \emph{et~al.}, {CompOSE Reference Manual}, Eur. Phys. J. A \textbf{58}
  (2022) 221.
\newblock \doi{10.1140/epja/s10050-022-00847-y}.
\newblock \eprint{2203.03209}.

\bibitem{Velten:2016bdk}
H.~Velten, A.~M. Oliveira and A.~Wojnar, {A free parametrized TOV: Modified
  Gravity from Newtonian to Relativistic Stars}, PoS \textbf{MPCS2015} (2016)
  025.
\newblock \doi{10.22323/1.262.0025}.
\newblock \eprint{1601.03000}.

\bibitem{Wagoner:1970vr}
R.~V. Wagoner, {Scalar tensor theory and gravitational waves}, Phys. Rev. D
  \textbf{1} (1970) 3209--3216.
\newblock \doi{10.1103/PhysRevD.1.3209}.

\bibitem{WagonerMalone}
R.~V. {Wagoner} and R.~C. {Malone}, {Post-Newtonian Neutron Stars},
  Astrophys.~J.~Lett. \textbf{189} (1974) L75.
\newblock \doi{10.1086/181468}.

\bibitem{Wex:2014nva}
N.~Wex.
\newblock {Testing Relativistic Gravity with Radio Pulsars} (2014).
\newblock \eprint{1402.5594}.

\bibitem{Will:1993hxu}
C.~M. Will.
\newblock {Theory and Experiment in Gravitational Physics} (1993).
\newblock ISBN 978-0-511-56424-6, 978-0-521-43973-2.
\newblock \doi{10.1017/CBO9780511564246}.

\bibitem{PW:book}
C.~M. Will.
\newblock {Gravity: Newtonian, Post-Newtonian, Relativistic}.
\newblock Cambridge University Press (2014).

\bibitem{Will:2014kxa}
C.~M. Will, {The Confrontation between General Relativity and Experiment},
  Living Rev. Rel. \textbf{17} (2014) 4.
\newblock \doi{10.12942/lrr-2014-4}.
\newblock \eprint{1403.7377}.

\bibitem{Wiringa:1988tp}
R.~B. Wiringa, V.~Fiks and A.~Fabrocini, {Equation of state for dense nucleon
  matter}, Phys. Rev. C \textbf{38} (1988) 1010--1037.
\newblock \doi{10.1103/PhysRevC.38.1010}.

\bibitem{Woodard:2015zca}
R.~P. Woodard, {Ostrogradsky's theorem on Hamiltonian instability},
  Scholarpedia \textbf{10} (2015) 32243.
\newblock \doi{10.4249/scholarpedia.32243}.
\newblock \eprint{1506.02210}.

\bibitem{Xu:2020vbs}
R.~Xu, Y.~Gao and L.~Shao, {Strong-field effects in massive scalar-tensor
  gravity for slowly spinning neutron stars and application to X-ray pulsar
  pulse profiles}, Phys. Rev. D \textbf{102} (2020) 064057.
\newblock \doi{10.1103/PhysRevD.102.064057}.
\newblock \eprint{2007.10080}.

\bibitem{Yagi:2012gp}
K.~Yagi, {A New constraint on scalar Gauss-Bonnet gravity and a possible
  explanation for the excess of the orbital decay rate in a low-mass X-ray
  binary}, Phys. Rev. D \textbf{86} (2012) 081504.
\newblock \doi{10.1103/PhysRevD.86.081504}.
\newblock \eprint{1204.4524}.

\bibitem{Yagi:2013ava}
K.~Yagi, D.~Blas, E.~Barausse \emph{et~al.}, {Constraints on Einstein-\AE{}ther
  theory and Ho\v{r}ava gravity from binary pulsar observations}, Phys. Rev. D
  \textbf{89} (2014) 084067.
\newblock \doi{10.1103/PhysRevD.89.084067}.
\newblock [Erratum: Phys.Rev.D 90, 069902 (2014), Erratum: Phys.Rev.D 90,
  069901 (2014)], \eprint{1311.7144}.

\bibitem{Yagi:2015oca}
K.~Yagi, L.~C. Stein and N.~Yunes, {Challenging the Presence of Scalar Charge
  and Dipolar Radiation in Binary Pulsars}, Phys. Rev. D \textbf{93} (2016)
  024010.
\newblock \doi{10.1103/PhysRevD.93.024010}.
\newblock \eprint{1510.02152}.

\bibitem{Yagi:2011xp}
K.~Yagi, L.~C. Stein, N.~Yunes \emph{et~al.}, {Post-Newtonian, Quasi-Circular
  Binary Inspirals in Quadratic Modified Gravity}, Phys. Rev. D \textbf{85}
  (2012) 064022.
\newblock \doi{10.1103/PhysRevD.85.064022}.
\newblock [Erratum: Phys.Rev.D 93, 029902 (2016)], \eprint{1110.5950}.

\bibitem{Yagi:2013mbt}
K.~Yagi, L.~C. Stein, N.~Yunes \emph{et~al.}, {Isolated and Binary Neutron
  Stars in Dynamical Chern-Simons Gravity}, Phys. Rev. D \textbf{87} (2013)
  084058.
\newblock \doi{10.1103/PhysRevD.87.084058}.
\newblock [Erratum: Phys.Rev.D 93, 089909 (2016)], \eprint{1302.1918}.

\bibitem{Yagi:2021loe}
K.~Yagi and M.~Stepniczka, {Neutron stars in scalar-tensor theories: Analytic
  scalar charges and universal relations}, Phys. Rev. D \textbf{104} (2021)
  044017.
\newblock \doi{10.1103/PhysRevD.104.044017}.
\newblock \eprint{2105.01614}.

\bibitem{Yagi:2013bca}
K.~Yagi and N.~Yunes, {I-Love-Q}, Science \textbf{341} (2013) 365--368.
\newblock \doi{10.1126/science.1236462}.
\newblock \eprint{1302.4499}.

\bibitem{Yagi:2013awa}
K.~Yagi and N.~Yunes, {I-Love-Q Relations in Neutron Stars and their
  Applications to Astrophysics, Gravitational Waves and Fundamental Physics},
  Phys. Rev. D \textbf{88} (2013) 023009.
\newblock \doi{10.1103/PhysRevD.88.023009}.
\newblock \eprint{1303.1528}.

\bibitem{Yagi:2016bkt}
K.~Yagi and N.~Yunes, {Approximate Universal Relations for Neutron Stars and
  Quark Stars}, Phys. Rept. \textbf{681} (2017) 1--72.
\newblock \doi{10.1016/j.physrep.2017.03.002}.
\newblock \eprint{1608.02582}.

\bibitem{Yunes:2009ke}
N.~Yunes and F.~Pretorius, {Fundamental Theoretical Bias in Gravitational Wave
  Astrophysics and the Parameterized Post-Einsteinian Framework}, Phys. Rev. D
  \textbf{80} (2009) 122003.
\newblock \doi{10.1103/PhysRevD.80.122003}.
\newblock \eprint{0909.3328}.

\bibitem{Yunes:2009ch}
N.~Yunes, D.~Psaltis, F.~Ozel \emph{et~al.}, {Constraining Parity Violation in
  Gravity with Measurements of Neutron-Star Moments of Inertia}, Phys. Rev. D
  \textbf{81} (2010) 064020.
\newblock \doi{10.1103/PhysRevD.81.064020}.
\newblock \eprint{0912.2736}.

\bibitem{Yunes:2016jcc}
N.~Yunes, K.~Yagi and F.~Pretorius, {Theoretical Physics Implications of the
  Binary Black-Hole Mergers GW150914 and GW151226}, Phys. Rev. D \textbf{94}
  (2016) 084002.
\newblock \doi{10.1103/PhysRevD.94.084002}.
\newblock \eprint{1603.08955}.

\bibitem{Zhao:2022vig}
J.~Zhao, P.~C.~C. Freire, M.~Kramer \emph{et~al.}, {Closing a
  spontaneous-scalarization window with binary pulsars}, Class. Quant. Grav.
  \textbf{39} (2022) 11LT01.
\newblock \doi{10.1088/1361-6382/ac69a3}.
\newblock \eprint{2201.03771}.

\end{thebibliography}

\end{document}